\documentclass[12pt,preprint]{aastex}
\usepackage{apjfonts}
\usepackage{epsfig}
\usepackage{amsmath}
\usepackage{graphicx}
\usepackage{color}
\usepackage{float}

\def\gsim{\lower 2pt \hbox{$\, \buildrel {\scriptstyle >}\over
{\scriptstyle \sim}\,$}}
\def\lsim{\lower 2pt \hbox{$\, \buildrel {\scriptstyle <}\over
{\scriptstyle \sim}\,$}}

\def\chandra{{\sl Chandra~}}
\def\xmm{{\sl XMM-Newton~}}
\def\hi{H~{I}}
\def\oi{O~{\scriptsize I}}
\def\oii{O~{\scriptsize II}}
\def\oiii{O~{\scriptsize III}}
\def\oiv{O~{\scriptsize IV}}
\def\ov{O~{\scriptsize V}}
\def\ovi{O~{\scriptsize VI}}
\def\ovii{O~{\scriptsize VII}}
\def\oviii{O~{\scriptsize VIII}}
\def\nei{Ne~{\scriptsize I}}
\def\neii{Ne~{\scriptsize II}}
\def\neiii{Ne~{\scriptsize III}}
\def\neiv{Ne~{\scriptsize IV}}
\def\nev{Ne~{\scriptsize V}}

\def\nevii{Ne~{\scriptsize VII}}

\def\neix{Ne~{\scriptsize IX}}
\def\nex{Ne~{\scriptsize X}}
\def\mgiii{Mg~{\scriptsize III}}
\def\mgiv{Mg~{\scriptsize IV}}
\def\mgv{Mg~{\scriptsize V}}
\def\mgxi{Mg~{\scriptsize XI}}
\def\fexvii{Fe~{\scriptsize XVII}}

\begin{document}

\title{WAVELENGTH MEASUREMENTS OF K TRANSITIONS OF OXYGEN, NEON, AND MAGNESIUM WITH X-RAY ABSORPTION LINES}

\author{Jin-Yuan Liao\altaffilmark{1}, Shuang-Nan Zhang\altaffilmark{1,2}, Yangsen Yao\altaffilmark{3,4}}
\altaffiltext{1}{Key Laboratory of Particle Astrophysics, Institute of High Energy Physics,
Chinese Academy of Sciences, Beijing 100049, China; zhangsn@ihep.ac.cn}
\altaffiltext{2}{National Astronomical Observatories, Chinese Academy of Sciences, Beijing, 100012, China}
\altaffiltext{3}{Eureka Scientific, 2452 Delmer Street Suite 100, Oakland, CA 94602}
\altaffiltext{4}{Center for Astrophysics and Space Astronomy, University of Colorado, Boulder, CO 80309, USA}
%\end{CJK*}

\begin{abstract}
Accurate atomic transition data are important in many astronomical research areas especially in line
spectroscopy study. Whereas transition data of He-like and H-like ions (i.e., ions at high-charge states)
are accurately calculated, that of K transitions of neutral or low-ionized metal elements are still
very uncertain. Spectroscopy of absorption lines produced in the interstellar medium (ISM) has been proven
to be an effective way to measure the central wavelengths of these atomic transitions. In this work we
analyze 36 {\sl Chandra} {\sl High Energy Transmission Grating} observations and search for and measure the ISM
absorption lines along sight lines of 11 low-mass X-ray binaries. We correct the Galactic rotation velocity to the rest frame for every observation and
then use two different methods to merge all the corrected spectra to a co-added spectrum.
However the co-added spectra obtained by these methods exhibit
biases, either to the observations of high counts or high signal-to-noise ratios of the lines.
We make Bayesian analysis to several significantly detected lines to obtain the systematic uncertainty and the bias correction of other lines. Compared to previous studies (e.g., Yao et al. 2009), our results improve the accuracy of wavelengths by a factor from two to five and
significantly reduce the systematic uncertainties and biases. Several weak transitions
(e.g., 1s--2p of \mgiv\ and \mgv; 1s--3p of \mgiii\ and \mgv) are also detected for the first time, albeit
with low significance; future observations with improved accuracy are required to confirm these detections.
\end{abstract}


\section{INTRODUCTION}           %% big picture of line physics, and the importance of line accuracy
Since the launch of modern X-ray space telescopes like \chandra and \xmm X-ray Observatories, X-ray astronomy
has entered the epoch of grating observations that can produce spectra with much improved energy resolution. In
these spectra narrow absorption/emission lines, which have never been observed before, now have been commonly
detected. These X-ray absorption/emission lines can be generated in a variety of astronomical environments,
e.g., stellar coronae (SC), supernova remnants, X-ray binaries (XRBs), galaxies, AGN, and interstellar media
(ISM) and intergalactic media (IGM) (e.g., \citealt{2003A&A...400.671}; \citealt{2008ApJ...676.L131};
\citealt{2002ApJ...565.1141}; \citealt{2006Nature...441.953}; \citealt{2003A&A...402.477}; \citealt{2001ApJ...554.L13};
\citealt{2005ApJ...624.751}; \citealt{2003ApJ...586.L49}; \citealt{2005ApJ...629.700}).
These lines carry valuable information about the absorbing/emitting material and thus are powerful diagnostic
tools for stellar evolution, explosion mechanism of SNe, mass exchange in accretion systems, interplays of
different galactic components, feedback of AGN, galaxy formation and evolution, evolution of our universe,
and so on. Clearly, proper identifications of these lines and subsequent scientific derivation and interpretation
strongly rely on the accuracy of atomic databases of responsible transitions.

There are several available databases of atomic transitions, among which the most commonly referenced four in
X-ray community are Verner et al. (1996, hereafter V96), NIST, XSTAR (\citealt{2004ApJS...155.675}), and APED
(\citealt{2001ApJ...556.L91}). While atomic data for ions at high-charge states (e.g., He- and H-like ions)
are very accurate in these databases and consistent with observations (Juett et al. 2004, 2006, J0406 hereafter; Yao et al. 2009, Y09 hereafter),
those for K transitions of neutral and mildly ionized metal elements (e.g., \oi--\oiii, \nei--\neiii, \mgiii--\mgv, etc.)
have not been included in any of them. There also exists other serious problems. First, although the statistical
errors of the wavelengths of these high-ionized lines can be less than 10 m\AA\ in astronomical observations,
they are still not enough for the study of the low-velocity gases. Y09 have measured these high-ionized lines;
however, some lines have large uncertainties, such as \oviii~K$\beta$ that has a statistical error 6.7 m\AA,
equivalent to 125 $\rm km~s^{-1}$. In addition, some lines (e.g., \nex~K$\alpha$) are too weak to be given
statistical errors. Second, the methods commonly used (e.g., Y09) to obtain the co-added spectrum introduce biases,
which must be corrected and otherwise will result in serious estimation errors of the gas velocities as well as
other parameters associated with the gas velocities. Finally, some of the line wavelengths are not consistent
with each other between the commonly referenced databases.
For example, the wavelength of \ovii~K$\beta$ in NIST is
18.6270 \AA\ but is 18.6288 \AA\ in V96. The difference between the two values is $\sim30$ $\rm km~s^{-1}$, which
is a serious problem for the study of the low-velocity gas. Therefore, it is essential to make observational
identifications of the K-shell transitions of the neutral, low-ionized, and high-ionized metals.
Since most of K transitions of these
low-ionization ions are in the wavelength range of 9.5--24.0 \AA, within which copious lines of highly ionized
metal elements have also been observed, these missing atomic data are important not only for their own rights
but also for proper identifications of other lines. Recently, several groups have calculated and updated these
databases (e.g., \citealt{gor00}; \citealt{2002ApJ...570.165}; \citealt{gor05}; \citealt{gar05}), but even the
centroid wavelengths, the most basic parameters of these lines, are yet to be coincided (see Table 1). In addition,
the laboratory measurements are also very uncertain; e.g., the wavelength of \oi\ 1s--2p given by Stolte et al. (1997)
range from 23.489 to 23.536 \AA.

%\begin{center}
%\renewcommand{\arraystretch}{1.2}
\begin{deluxetable}{lccccc}
  %\tabletypesize{\scriptsize}
  \tablewidth{0pt}
  \tablecaption{The wavelengths of low-ionized O, Ne, and Mg from previous work}
  \tablehead{  &            & \multicolumn{4}{c}{Wavelength (\AA)} \\
                                         \cline{3-6}
      Ion      & Transition & G05BN02   & G00GM05     & J0406     & Y09
            }
  \startdata
      \oi\     & 1s--2p     & $23.4475$ & $23.532$    & $23.508$  & $23.508$ \\%% done
      \oi\     & 1s--3p     & $\cdots$  & $22.907$    & $22.884$  & $\cdots$ \\%% done
      \oii\    & 1s--2p     & $23.3100$ & $22.781$    & $23.330$  & $23.384$ \\%% done
      \oii\    & 1s--3p     & $\cdots$  & $22.576$    & $\cdots$  & $\cdots$ \\%% done
      \oiii\   & 1s--2p     & $23.0800$ & $\cdots$    & $23.140$  & $\cdots$ \\%% done
      \oiii\   & 1s--3p     & $\cdots$  & $\cdots$    & $\cdots$  & $\cdots$ \\%% done
      \hline
      \nei\    & 1s--3p     & $\cdots$  & $14.295$    & $14.295$  & $14.294$ \\%% done
      \neii\   & 1s--2p     & $14.6310$ & $14.608$    & $14.608$  & $14.605$ \\%% done
      \neii\   & 1s--3p     & $\cdots$  & $\cdots$    & $\cdots$  & $14.001$ \\%% done
      \neiii\  & 1s--2p     & $14.5260$ & $14.508$    & $14.508$  & $14.507$ \\%% done
      \neiii\  & 1s--3p     & $\cdots$  & $\cdots$    & $\cdots$  & $13.690$ \\%% done
      \hline
      \mgiii\  & 1s--3p     & $\cdots$  & $\cdots$    & $\cdots$  & $\cdots$ \\ %% done
      \mgiv\   & 1s--2p     & $\cdots$  & $\cdots$    & $\cdots$  & $\cdots$ \\ %% done
      \mgiv\   & 1s--3p     & $\cdots$  & $\cdots$    & $\cdots$  & $\cdots$
      \enddata
    \tablerefs{G05BN02: Garc\'ia et al.(2005) and Behar \& Netzer (2002);
               G00GM05: Gorczyca (2000) and Gorczyca \& McLaughlin (2005);
               J0406: Juett et al. (2004, 2006); Y09: Yao et al. (2009).}
\end{deluxetable}
%\end{center}

Some Galactic X-ray sources are very bright in the soft X-ray band.
When the X-ray continuous radiations pass through the ISM in different phases,
X-ray absorption lines of ions at various charge states are produced and thus are
excellent calibration references for these atomic data.
For instance, Schattenburg \& Canizares (1986) used the Einstein observations of the
Crab to obtain the wavelength of the neutral O (1s--2p of \oi), although the result is
uncertain ($23.46\pm0.22$ \AA) according to modern standard.
J0406 analyzed {\sl Chandra} High Energy Transmission Grating (HETG)
observations of several Galactic XRBs and obtained the wavelengths of some strong lines
(i.e., 1s--2p of \oi, \oii, \oiii, \neii, \neiii\ and 1s--3p of \oi).
They found theoretically calculated line centroids of K transitions of low
ionized neon and oxygen need to be shifted $>20$ m\AA\ to match observed
values, which corresponds to $300-400~{\rm km~s^{-1}}$ shift for
oxygen- and neon-absorbers.
However, the ISM is rotating around the Galactic
center and thus the observed X-ray lines are not at the rest-frame wavelengths. Unfortunately,
the above three previous works did not do the Galactic rotation correction.
It is also necessary to exclude X-ray sources with significant intrinsic absorptions, such as accreting X-ray binaries
with winds from either their companions or accretion disks along the line of sight (LOS). This can be done easily by comparing absorption
line and/or total absorption properties of a target with multiple observations, since all X-ray binaries exhibit significant variability.

Although X-ray absorption lines produced in the ISM are indeed ideal
sources for calibrating theoretical predictions of atomic transitions,
high spectral quality is crucial to accomplishing this important task.
The measurements by J0406 are based on relatively poor quality
spectra (with signal-to-noise ratio SNR$\lsim15$ per 10-m\AA\
spectral bin) obtained with short observations of several sources; thus
only those strong transitions (e.g., 1s--2p of \oi, \neii, and \neiii) were
relatively well constrained, and others were still of large uncertainties
($\Delta\lambda\sim6-20$ m\AA) or not observed.
Recently Y09 presented an extensive study of ISM X-ray
absorption lines in the spectrum of Cyg~X--2 observed with the
{\sl Chandra}-HETG spectrograph.
The high spectral quality not only allowed them to
measure most of the transitions listed by J0406 as
accurate as $\Delta\lambda\sim1-4$ m\AA\ (1$\sigma$ uncertainty) but
also enabled them to constrain other faint
transitions not included and/or misidentified in their list.
For instance, Y09 detected and measured the 1s--3p transition of \neiii\
that could not be revealed in previous poor quality spectra, and they did
not confirm the reported line at 23.140 \AA\ that was misidentified
as 1s--3p transition of \oiii.

In this work, our aim is to obtain more accurate K transition data of
neutral, low- and high-ionized metals, as well as to find
some weak absorption lines (e.g., 1s--2p of \mgiv--\mgv; 1s--3p of \mgiii--\mgv).
We jointly analyze 36
{\sl Chandra}-HETG observations of 11 Galactic XRBs and give the most accurate
wavelengths for K-shell transitions of neutral, low- and
high-ionized atoms.
In Section 2 we describe our methods of extracting the X-ray spectra and correcting for the Galactic rotation.
In Section 3 we present the results of a joint analysis of all the observational data.
In Section 4 we estimate systematic uncertainties and make necessary corrections to the detected lines.
A discussion and summary are given in Section 5 and 6, respectively.

\section{SAMPLE SELECTION, DATA PROCESSING AND GALACTIC ROTATION CORRECTION}
\subsection{Sample Selection}
  In order to obtain high SNR continuum spectra and absorption lines, we adopt the
following criteria to select the sample:
\begin{itemize}
\item Only the Galactic low-mass X-ray binaries (LMXBs) are used as the background light sources. Compared to
extragalactic sources (e.g., AGNs), Galactic XRBs are usually located at low Galactic latitudes and thus the
column densities of the ISM along the LOS are usually large if they are not too close to us.
Thus strong absorption lines are expected in their spectra. High mass X-ray binaries are excluded to avoid any
possible contamination caused by their stellar winds that may not be stationary (e.g., Cyg X--1;
\citealt{2002ApJ...565.1141});
\item Sources with intrinsic absorptions or emissions (e.g., GX 339--4, \citealt{2004ApJ...601.450}; 4U 1916--05,
\citealt{2006ApJ...646.493}) are excluded;
\item Only sources with Galactic latitudes more than 2 degrees are chosen to avoid too heavy Galactic absorptions.
Most lines we concern are at wavelengths above 10 \AA\ and the photons in this band suffer heavy absorption in the Galactic plane,
which can reduce the significance of the line fitting (Appendix A);
\item Only HETG observations are used to assure a high spectral resolution;
\item In order to ensure that there are significant absorption lines, we only include the sources for which the
strongest O line (1s--2p of \oi) and the strongest Ne line (1s--2p of \neii) (Figures 1 and 2) are detected in
the pipe-line-produced spectra. We choose \neii~K$\alpha$ as an indicator, although it is always intrinsically
weaker than \oi~K$\alpha$ lines; the latter are not detected from several sources due to the rapidly decreasing
effective area of the HETG at wavelengths longer than 20 \AA.
\end{itemize}

  Finally, 36 {\sl Chandra}-HETG observations of 11 LMXBs are selected as listed in Table 2.

\begin{figure}
\center{
\includegraphics[angle=0,scale=1.0]{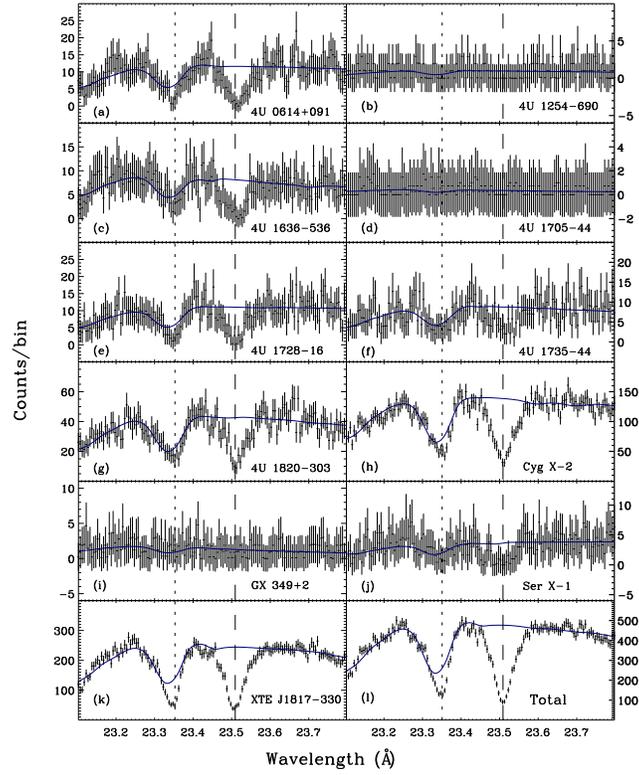}}
\caption{Panels (a) to (k) present the positions of the absorption lines of \oi~K$\alpha$ (dashed line) and \oii~K$\alpha$ (dotted line)
         of the targets we selected to study. The co-added spectrum of all the 11 targets is shown in panel (l).}
\label{Fig:1}
\end{figure}

\begin{figure}
\center{
\includegraphics[angle=0,scale=1.0]{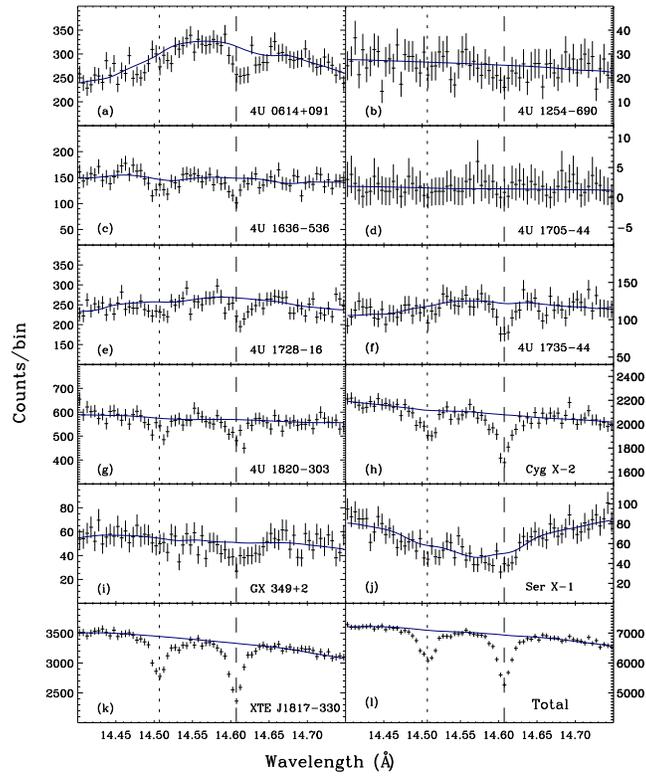}}
\caption{The same as Figure 1, but the positions of the absorption lines of \neii~K$\alpha$ (dashed line) and \neiii~K$\alpha$ (dotted line) are shown.}
\label{Fig:2}
\end{figure}

%\begin{center}
\begin{tiny}
\begin{deluxetable}{cccccccc}
	\tabletypesize{\scriptsize}
  \tablewidth{0pt}
  \tablecaption{Targets in our sample}
  \tablehead{    Src.name    &     ($l, b$)     &  Obs.ID  & EXPT & ACR  &  Distance  &  $V_{\rm l}$  &  $V_{\rm h}$  \\
                             &                  &  & (ks)  & (count  s$^{-1}$) & (kpc) & ($\rm km~s^{-1}$) & ($\rm km~s^{-1}$)
            }
  \startdata
                4U 1705--44  & (343.32, -2.36)  & 1923   &  25  &  117  & $5.8^a$  &  -17     &  -37      \\%% done
                             &                  & 1924   &  6   &  171  & $\cdots$ & $\cdots$ & $\cdots$  \\%% done
                             &                  & 5500   &  27  &  37   & $\cdots$ & $\cdots$ & $\cdots$  \\%% done
                \cline{3-8}
                GX 349+2     & (349.10, 2.75)   & 715    &  11  &  267  & $5.0^c$  &  -7      &  -10      \\%% done
                             &                  & 3354   &  35  &  281  & $\cdots$ & $\cdots$ & $\cdots$  \\%% done
                             &                  & 6628   &  13  &  301  & $\cdots$ & $\cdots$ & $\cdots$  \\%% done
                             &                  & 7336   &  12  &  264  & $\cdots$ & $\cdots$ & $\cdots$  \\%% done
                \cline{3-8}
                4U 0614+091  & (200.88, -3.36)  & 10759  &  60  &  42   & $3.2^b$  &  19      &  36       \\%% done
                             &                  & 10760  &  45  &  45   & $\cdots$ & $\cdots$ & $\cdots$  \\%% done
                             &                  & 10857  &  59  &  62   & $\cdots$ & $\cdots$ & $\cdots$  \\%% done
                             &                  & 10858  &  35  &  39   & $\cdots$ & $\cdots$ & $\cdots$  \\%% done
                \cline{3-8}
                4U 1636--536 & (332.92, -4.82)  & 105    &  30  &  105  & $5.95^a$ &  -14     &  -45      \\%% done
                             &                  & 1939   &  27  &  93   & $\cdots$ & $\cdots$ & $\cdots$  \\%% done
                             &                  & 6635   &  23  &  41   & $\cdots$ & $\cdots$ & $\cdots$  \\%% done
                             &                  & 6636   &  25  &  102  & $\cdots$ & $\cdots$ & $\cdots$  \\%% done
                \cline{3-8}
                Ser X--1     & (36.12, 4.84)    & 700    &  78  &  110  & $7.7^a$  &  14      &  57       \\%% done
                \cline{3-8}
                4U 1254--690 & (303.48, -6.42)  & 3823   &  53  &  27   & $15.5^a$ &  -2      &  15       \\%% done
                \cline{3-8}
                4U 1735--44  & (346.05, -6.70)  & 704    &  25  &  92   & $6.5^a$  &  -3      &  -8       \\%% done
                             &                  & 6637   &  24  &  103  & $\cdots$ & $\cdots$ & $\cdots$  \\%% done
                             &                  & 6638   &  23  &  110  & $\cdots$ & $\cdots$ & $\cdots$  \\%% done
                \cline{3-8}
                4U 1820--303  & (2.79, -7.91)    & 1021   &  10  &  124  & $4.94^a$ &  6       &  -1       \\%% done
                             &                  & 1022   &  11  &  142  & $\cdots$ & $\cdots$ & $\cdots$  \\%% done
                             &                  & 6633   &  25  &  215  & $\cdots$ & $\cdots$ & $\cdots$  \\%% done
                             &                  & 6634   &  25  &  272  & $\cdots$ & $\cdots$ & $\cdots$  \\%% done
                             &                  & 7032   &  46  &  242  & $\cdots$ & $\cdots$ & $\cdots$  \\%% done
                \cline{3-8}
              XTE J1817--330 & (359.82, -8.00)  & 6615   &  50  &  998  & $5.0^c$  &  7       &  0        \\%% done
                             &                  & 6616   &  50  &  560  & $\cdots$ & $\cdots$ & $\cdots$  \\%% done
                             &                  & 6617   &  47  &  293  & $\cdots$ & $\cdots$ & $\cdots$  \\%% done
                             &                  & 6618   &  51  &  109  & $\cdots$ & $\cdots$ & $\cdots$  \\%% done
                \cline{3-8}
                4U 1728--16  & (8.51, 9.04)     & 703    &  21  &  129  & $5.0^c$  &  7       &  -2       \\%% done
                             &                  & 11072  &  98  &  118  & $\cdots$ & $\cdots$ & $\cdots$  \\%% done
                \cline{3-8}
                 Cyg X--2    & (87.33, -11.32)  & 1016   &  15  &  280  & $11.0^a$ &  -8      &  -54      \\%% done
                             &                  & 1102   &  29  &  130  & $\cdots$ & $\cdots$ & $\cdots$  \\%% done
                             &                  & 8170   &  77  &  384  & $\cdots$ & $\cdots$ & $\cdots$  \\%% done
                             &                  & 8599   &  71  &  382  & $\cdots$ & $\cdots$ & $\cdots$  \\%% done
                             &                  & 10881  &  67  &  267  & $\cdots$ & $\cdots$ & $\cdots$
    \enddata
    \tablecomments{We obtained the distances of the targets from: a. Galloway et al. (2008); b. Kuulkers et al. (2010);
               c. We use the values from Kong (2006), Iaria et al. (2004) and Sala et al. (2007), since the distances of these
               sources have not been confirmed. `EXPT' in column 4 is the exposure time and `ACR' in column 5
               is the average counts rate of every Obs.ID. $V_{\rm l}$ and $V_{\rm h}$ in column 7 and 8 are the velocities relative to the LSR
               of the low-ionized and high-ionized gases respectively.}
\end{deluxetable}
\end{tiny}
%\end{center}

\subsection{Data Reduction}
  We analyze all observations using CIAO 4.4 and CALDB 4.4.7. We use the standard tool {\sl tgextract} to extract
the spectra, i.e. PHA files. The energy redistribution matrix file (RMF) and the ancillary response file (ARF) are
made by the standard tools {\sl mkgrmf} and {\sl fullgarf} respectively. All the steps follow the standard
procedures except for the determination of the position of the zeroth-order image of each source, which is the
key to fix the wavelength scale. Since all the targets in our sample are bright sources, the zeroth-order source
images are expected to be either severely piled-up in observations with the timed exposure (TE) mode or have been
compressed onto several pixels in observations with the continuous clocking (CC) mode. Rather than following the
standard script to find the source positions, we use the mean position of the crosses between the CCD read-out
streaks and each of the two arms (HEG and MEG) as the source positions for TE-mode observations, and use two
Gaussian profiles to fit the compressed image to determine the source positions for CC-mode observations.

\subsection{Velocity Correction due to the Galactic Rotation}
  Our purpose is to obtain the wavelengths of the absorption lines accurately and thus
we must correct for the fact that the line-absorbing gas is not at rest with respect
to the local standard rest frame (LSR). The gas in the Galactic plane is
rotating around the Galactic center and the rotation velocity increases as the radius decreases. The absorption
lines in X-ray band are produced by the multi-phase ISM along the LOS, and the motions are also different for the
ISM in different phases. Therefore, we must correct the Galactic rotation of neutral, low-ionized gas, and
high-ionized gas separately. We adopt the method in Y09 that assumes that the gas is rotating around the Galactic
center approximately in circular orbit and the closer to the Galactic center the faster gas rotates. For gas in
radius $R$ and with a rotational speed $V$, the velocity relative to the LSR is
\begin{equation}
V_r = V_{r,\ {\rm gas}}-V_{r,\ {\rm sun}} = R_0 \cos b \sin l (\frac{V}{R}-\frac{V_0}{R_0}),
\end{equation}
where
\begin{equation}
R = \sqrt{R_0^2 + D^2\cos^2 b + 2DR_0\cos b \cos l},
\end{equation}
here $D$ is the distance between the gas and the observer, $R_0$ and $V_0$ are the radius of the LSR and its rotation velocity, respectively (\citealt{spa00}). The
average velocity can be obtained by integrating all the gas with different $R$ and $V$ along the path to the source.

  For low-ionized gas, we use the neutral \hi\ 21cm emission to trace the gas velocity. Data are derived from the
Leiden Argentine Bonn (LAB) Galactic \hi\ Survey (\citealt{2005A&A...440.775}) that has a $\sim30'$ spatial grid
and 1.3 $\rm km~s^{-1}$ velocity resolution. To obtain the average profile ($\phi_{\rm m}$) of the \hi\ 21cm
emission toward the source, we average the emission profiles from the four adjacent \hi\ observations with
respect to their angular separations, i.e.,
\begin{equation}
\phi_{\rm m} = \frac{\sum_{i=1}^4 \phi_i / d_i^2}{\sum_{i=1}^4 1 / d_i^2},
\end{equation}
where $\phi_i$ is the profile of the \hi\ 21cm emission along each individual LOS and $d_i$ is the angular
separation between the LOS and the source. Then we can obtain the average \hi\ velocity from $\phi_{\rm m}$ and
use this value as the projected velocity of the neutral or low-ionized gas along the LOS.

  For high-ionized gas, as its scale height is much larger than the low-ionized gas, the halo-lagging effect (Rand 1997, 2000)
must be considered. Here, we follow the three assumptions in Y09: (1) the density of the hot gas is $n = n_0 e^{-z/z_n}$,
where $n_0$ is the gas density in the Galactic plane and $z$ is the height from the Galactic plane and $z_n$ = 3 kpc
(e.g., \citealt{2008ApJS...176.59}; \citealt{2008ApJ...672.L21}); (2) the rotation velocity linearly decreases from $V_0$
to zero at a height $z_0$ = 8 kpc (Rand 1997) above the Galactic disk; (3) the velocity at radii between $R_0$ and $R$ can be linearly
interpolated from $V_0$ and $V$. For the LSR, we take $R_0$ = 8 kpc and $V_0$ = 220 $\rm km~s^{-1}$. The rotation velocity
near the target can be inferred from the smallest (largest) velocity of the \hi\ emission and then the average velocity
of the hot gas can be obtained by integrating  all the gas with different $R$ and $V$ in LOS. The average velocities of
the neutral or low-ionized gas and high-ionized hot gas are listed in Table 2.

  We transform all the spectra to the LSR and then use the IDL scripts {\sl writepha} and {\sl wrt\_ogip\_rmf} to produce
the spectrum file (PHA) and the response file (RSP = RMF $\times$ ARF) respectively.

\begin{figure*}[ht]
\center{
\includegraphics[angle=0,scale=1.0]{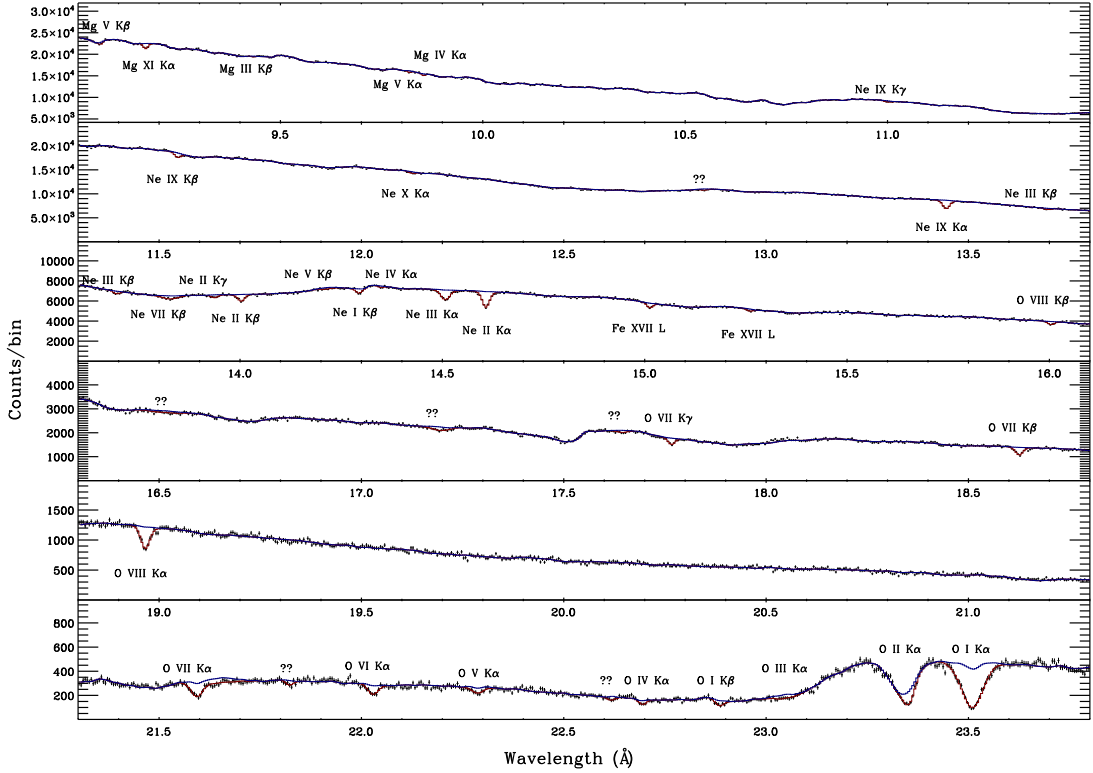}}
\caption{Co-added MEG spectrum with the best-fit continuum (thick blue lines). Red histograms mark the observed absorption lines
         and question marks indicate those unidentified ones.}
\label{Fig:3}
\end{figure*}

\begin{center}
\begin{figure*}[ht]
\center{
\includegraphics[angle=0,scale=1.0]{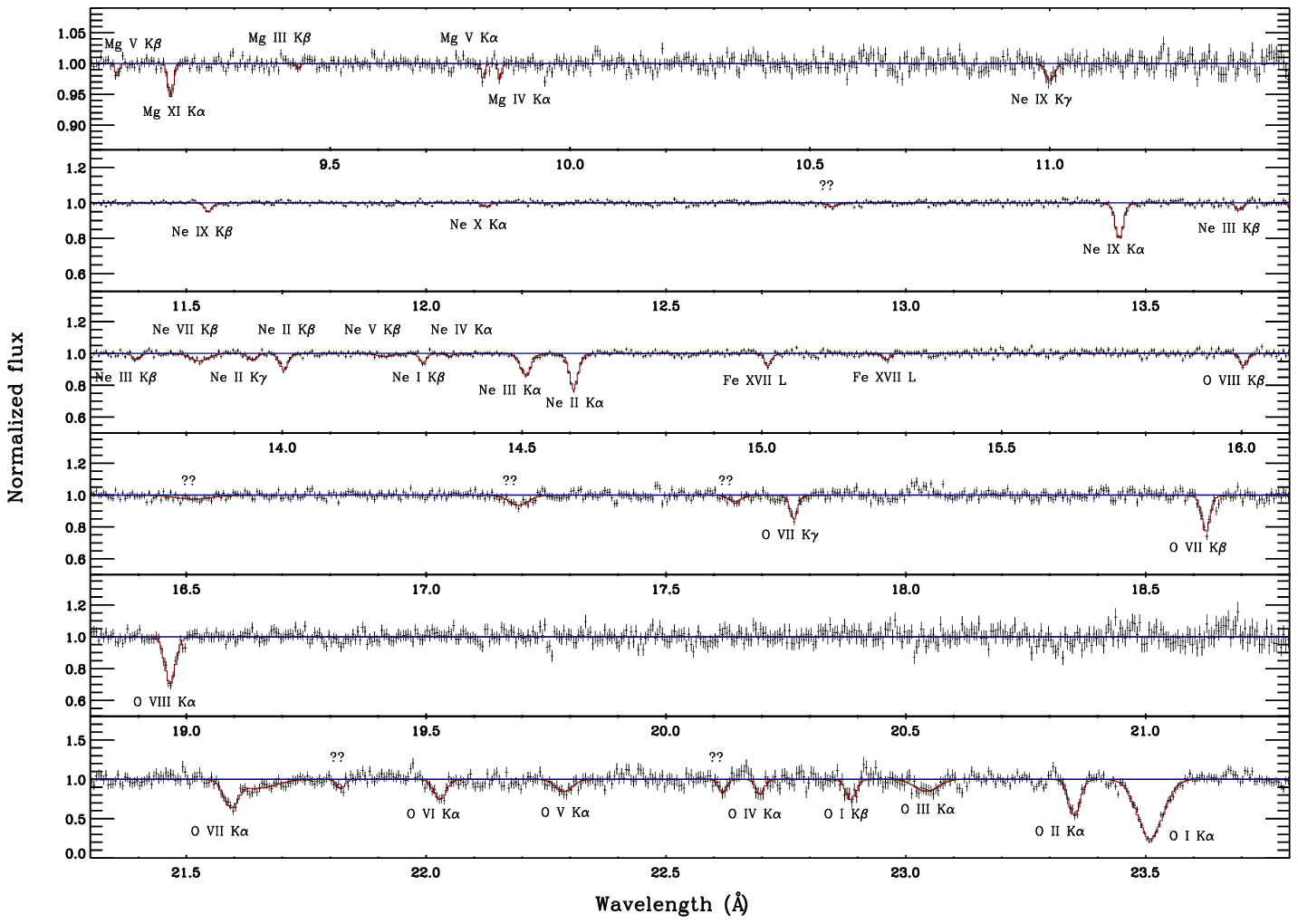}}
\caption{The same as Figure 1, but the spectrum is normalized to the best-fit continuum.}
\label{Fig:4}
\end{figure*}
\end{center}

\section{MERGING SPECTRA AND DETERMINATION OF THE LINE CENTRAL WAVELENGTHS}
  In this section, we use two different methods to jointly analyze these 36 observations to obtain the wavelengths of
the absorption lines of neutral, low-ionized, moderate-ionized and high-ionized gas (Table 4). The result and discussion
will be presented in Sections 5 and 6.

\subsection{Method 1: Direct Merging of All Full Spectra}
  The entire 36 spectra are co-added by the following two steps to increase the SNR as the method described in Y09:
(1) add the counts of each spectrum channel by channel to make a new spectrum file (PHA); (2) merge all the response
files to Producing the response file (RSP), where we use the total counts as the weights of each observation.

  We fit the co-added spectrum with XSPEC (version 12.7.0) to analyze the co-added spectrum. The continuum is fitted
using a power-law plus several broad Gaussian functions and the absorption line is fitted by a narrow Gaussian function,
\begin{equation}
\phi(\lambda_i) = a \frac{1}{\sqrt{2 \pi} b} e^{\frac{(\lambda_i-\lambda)^2}{2b^2}},
\end{equation}
where $\lambda$ is the wavelength of the line centroid, $b$ is the width of the line and $a$ is the normalization of the
line. From Figures 3 and 4, it can be seen that the co-added spectrum has very high SNR. As a result, many weak lines
(e.g., \ovii~K$\gamma$) are visible. The SNRs around the \oi~K$\alpha$, \neii~K$\alpha$, and \mgxi~K$\alpha$ lines are
about 20, 80, 110 respectively.

  Merging the spectra by this method can generally increase the SNR significantly. However, the spectra with weak lines
may also weaken the SNR. We take two observations (ObsID 6618 and 8599) as examples. The wavelength of \ovi~K$\alpha$
is $21.592_{-7}^{+14}$ \AA\ obtained from the ObsID 6618, whereas it is $21.597_{-10}^{+15}$ \AA\ from the co-added spectrum
with ObsID 6618 and 8599. The error actually increases as more observations are combined, since the significance of the
absorption line in ObsID 8599 is too low to increase the significance of the line in the co-added spectrum. As a result,
the spectrum with weak absorption lines can only increase the total counts of the continuum, which increases the background
noise to the line (Figure 5). To avoid this problem, another method of merging the spectra is investigated next.

\begin{figure}[!t]
\center{
\includegraphics[angle=0,scale=0.70]{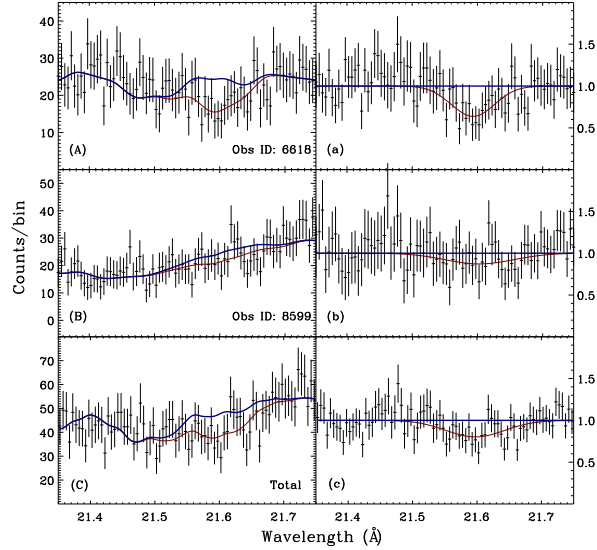}}
\caption{The \ovii~K$\alpha$ line in Obs-ID 6618, 8599, and the co-added spectrum. The right column is normalized to the best-fit continuum.}
\label{Fig:5}
\end{figure}

\subsection{Method 2: Weighted Merging of Net Lines}

  Unlike the method in Section 3.1, we fit the spectrum to obtain the continuum and line parameters
($\lambda$, $b$, and $a$) of each observation first. All data for each line flux are then merged
to produce a co-added line flux after the removal of the continuum of each individual spectrum,
\begin{equation}
f_{\rm add}=\frac{\sum_{i=1}^{36} f_i w_i}{\sum_{i=1}^{36} w_i},
\end{equation}
where $f_{\rm add}$ is the flux of each co-added line flux, $f_i$ and $w_i$ are the flux and weight for each line in
each observation. In order to avoid the problem caused by the spectrum with weak ISM absorption, we must evaluate $w_i$
carefully. Here, the weight of each absorption
line is given as (Appendix B)
\begin{equation}
w_i=\frac{a_i}{\sigma_i^2 b_i},
\end{equation}
where $\sigma_i$ is the error of the continuum around the line, $a_i$ and $b_i$ are the same as that defined in Equation (4).

For the strong lines
(e.g., \oi~K$\alpha$ and \neii~K$\alpha$), the significance is high enough to determine the line parameters for
most observations, which can be used as the weight to merge the spectra.
However, the significance of the weak lines (e.g., \oiii~K$\alpha$ and \neiii~K$\beta$) is too low and the line
parameters can be obtained only in few observations. Among all the 36 observations, $N($\neii~K$\alpha$; $SNR>1.645)=26$;
but $N($\neiii~K$\beta$; $SNR>1.645)=3$. To solve this problem, we make a simple assumption
that the clouds in every LOS have the same ion fraction (i.e., all the lines in each spectrum have the same
line-strength ratios). The ratios \oi~K$\alpha/$\oii~K$\alpha$, \oviii~K$\alpha/$\ovii~K$\beta$, and
\neii~K$\alpha/$\neiii~K$\alpha$ are shown in Figure 6, which proves the validity of the assumption.
As shown in Table 3, all the lines are divided into five groups and the strongest line of
each group is also found. Then we can use
$w_i$ of the five strongest lines of each group to merge the spectra to obtain the wavelength of each line in
each group. For the low-ionized O lines, $w_i$(\oi~K$\alpha$) is used as the weight.
$w_i$(\mgxi~K$\alpha$) is used as the weight to merge the spectra both for the low-ionized and the
high-ionized Mg absorption lines, since the \mgxi~K$\alpha$ line is the only Mg absorption line detected significantly.

The co-added spectrum also exhibits very high SNR and we do the same Gaussian fit to every line at
neutral, low-ionized, and high-ionized states. The results of neutral or low-ionized, moderate-ionized
and high-ionized lines are shown in Table 4.

\begin{figure}[!t]
\center{
\includegraphics[angle=0,scale=0.60]{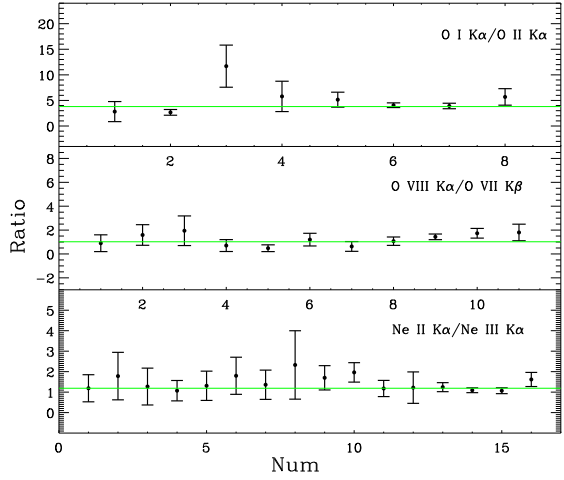}}
\caption{The ratio of line strengths between \oi~K$\alpha/$\oii~K$\alpha$, \oviii~K$\alpha/$\ovii~K$\beta$,
         as well as \neii~K$\alpha/$\neiii~K$\alpha$. The numbers of observations with both lines are well
         fitted are: $N$(\oi~K$\alpha/$\oii~K$\alpha)=8$, $N$(\oviii~K$\alpha/$\ovii~K$\beta)=11$, and
         $N$(\neii~K$\alpha/$\neiii~K$\alpha)=16$.}
\label{Fig:6}
\end{figure}

\begin{tiny}
\begin{deluxetable}{cp{2.0cm}p{2.0cm}p{2.2cm}p{2.2cm}p{2.2cm}}
  %\tabletypesize{\scriptsize}
  \tablewidth{0pt}
  \tablecaption{Five different line groups with different ionized states and different elements}
  \tablehead{           &  O (L)  &  O (M \& H)  & Ne (L) &  Ne (M \& H)  &  Mg (A)  }
  \startdata
             strongest  &  \oi~K$\alpha$  &  \oviii~K$\alpha$  &  \neii~K$\alpha$  &  \neix~K$\alpha$  &  \mgxi~K$\alpha$    \\%% done
             \hline
             1s--2p &  \oi, \oii, \oiii\ & \oiv, \ov, \ovi, \ovii, \oviii\ & \neii, \neiii\ & \neiv, \nev, \nevii, \neix, \nex\ & \mgiv, \mgv, \mgxi\ \\%% done
			 1s--3p &  \oi              & \ovii, \oviii\                     & \nei, \neii, \neiii\   & \neix\                          & \mgiii, \mgv\  \\%% done
			 1s--4p &                  &  \ovii\                               &   \neii\               &  \neix\                       &
  \enddata
  \tablecomments{The 1st row are the group names, where `L', `M', and `H' means the low-ionized, moderate-ionized, and high-ionized state.
                 `A' means all ionized states. The 2nd row lists the strongest lines in each group.}
\end{deluxetable}
\end{tiny}

%\begin{center}
%\renewcommand{\arraystretch}{1.2}
\begin{tiny}
\begin{deluxetable}{lccccc}
  \tabletypesize{\scriptsize}
  \tablewidth{0pt}
  \tablecaption{The wavelengths of low-ionized, moderate-ionized, and high-ionized elements}
  \tablehead{  Ion      & Transition  & $\lambda_1$(\AA)  & $snr_{1}$ & $\lambda_2$(\AA) & $snr_{2}$}
  \startdata
               \oi\     & 1s--2p     & $23.5087_{-0.6}^{+0.6}$   & $49.9$    & $23.5091_{-0.7}^{+0.7}$   & $44.1$ \\%% done
               \oi\     & 1s--3p     & $22.8834_{-2.3}^{+2.3}$   & $5.5 $    & $22.8872_{-2.3}^{+2.4}$   & $3.1 $ \\%% done
               \oii\    & 1s--2p     & $23.3508_{-0.9}^{+1.3}$   & $14.2$    & $23.3507_{-1.1}^{+1.1}$   & $13.4$ \\%% done
               \oiii\   & 1s--2p     & $23.0392_{-5.4}^{+8.5}$   & $4.2 $    & $23.0565_{-8.3}^{+7.1}$   & $2.4 $ \\%% done
               \oiv\    & 1s--2p     & $22.6969_{-37.3}^{+2.9}$  & $4.1 $    & $22.6967_{-3.3}^{+3.4}$   & $2.9 $ \\%% done
               \ov\     & 1s--2p     & $22.2849_{-3.3}^{+4.7}$   & $4.4 $    & $22.2872_{-4.2}^{+3.0}$   & $3.1 $ \\%% done
               \ovi\    & 1s--2p     & $22.0281_{-1.4}^{+2.5}$   & $6.3 $    & $22.0292_{-2.2}^{+1.8}$   & $7.4 $ \\%% done
		       \ovii\   & 1s--2p     & $21.5915_{-1.3}^{+1.6}$   & $7.1 $    & $21.5948_{-1.5}^{+1.4}$   & $10.6$ \\%% done
		       \ovii\   & 1s--3p     & $18.6259_{-0.7}^{+1.1}$   & $13.0$    & $18.6255_{-0.9}^{+1.1}$   & $10.9$ \\%% done
		       \ovii\   & 1s--4p     & $17.7657_{-1.2}^{+0.9}$   & $9.5 $    & $17.7673_{-1.3}^{+0.8}$   & $6.1 $ \\%% done
		       \oviii\  & 1s--2p     & $18.9667_{-0.9}^{+0.6}$   & $18.1$    & $18.9664_{-0.7}^{+0.8}$   & $16.3$ \\%% done
		       \oviii\  & 1s--3p     & $16.0046_{-1.3}^{+0.6}$   & $8.3 $    & $16.0044_{-1.3}^{+1.3}$   & $8.2 $ \\%% done
               \hline
               \nei\    & 1s--3p     & $14.2937_{-2.9}^{+0.9}$   & $7.8 $    & $14.2942_{-0.7}^{+0.7}$   & $10.7$ \\%% done
               \neii\   & 1s--2p     & $14.6068_{-0.3}^{+0.4}$   & $29.4$    & $14.6071_{-0.4}^{+0.4}$   & $28.0$ \\%% done
               \neii\   & 1s--3p     & $14.0029_{-0.7}^{+1.0}$   & $11.6$    & $14.0031_{-0.9}^{+0.9}$   & $11.3$ \\%% done
               \neii\   & 1s--4p     & $13.9373_{-3.0}^{+1.5}$   & $4.5 $    & $13.9339_{-2.9}^{+3.2}$   & $4.5 $ \\%% done
               \neiii\  & 1s--2p     & $14.5068_{-0.6}^{+0.7}$   & $15.0$    & $14.5073_{-0.7}^{+0.7}$   & $18.1$ \\%% done
               \neiii\  & 1s--3p     & $13.6951_{-2.4}^{+1.4}$   & $3.8 $    & $13.6953_{-2.5}^{+1.7}$   & $5.1 $ \\%% done
               \neiv\   & 1s--2p     & $14.3471_{-64.3}^{+64.3}$ & $1.8 $    & $\cdots$                  & $\cdots$ \\%% done
               \nev\    & 1s--2p     & $14.2127_{-38.6}^{+10.2}$ & $3.2 $    & $14.2090_{-13.7}^{+12.6}$ & $2.2 $ \\%% done
               \nevii\  & 1s--2p     & $13.8272_{-1.9}^{+4.2}$   & $5.3 $    & $13.8269_{-4.2}^{+3.3}$   & $3.2 $ \\%% done
		       \neix\   & 1s--2p     & $13.4455_{-0.3}^{+0.5}$   & $27.6$    & $13.4453_{-0.4}^{+0.4}$   & $26.1$ \\%% done
		       \neix\   & 1s--3p     & $11.5459_{-0.7}^{+1.4}$   & $11.0$    & $11.5456_{-1.0}^{+1.6}$   & $9.9 $ \\%% done
		       \neix\   & 1s--4p     & $11.0010_{-2.5}^{+2.5}$   & $4.1 $    & $10.9987_{-1.8}^{+1.9}$   & $2.6 $ \\%% done
		       \nex\    & 1s--2p     & $12.1250_{-4.1}^{+2.1}$   & $3.4 $    & $12.1253_{-2.4}^{+2.9}$   & $3.8 $ \\%% done
               \hline
               \mgiii\  & 1s--3p     & $9.4330_{-22.1}^{+44.8}$  & $1.5 $    & $9.4766_{-4.1}^{+0.6}$    & $3.2 $ \\ %% done
		       \mgiv\   & 1s--2p     & $9.8539_{-1.0}^{+29.1}$   & $1.5 $    & $\cdots$                  & $\cdots$ \\ %% done
               \mgv\    & 1s--2p     & $9.8191_{-30.1}^{+30.1}$  & $3.3 $    & $9.8195_{-1.9}^{+1.4}$    & $1.0 $ \\ %% done
               \mgv\    & 1s--3p     & $9.0569_{-2.0}^{+28.0}$   & $3.0 $    & $9.0553_{-1.9}^{+2.5}$    & $2.3 $ \\ %% done
		       \mgxi\   & 1s--2p     & $9.1679_{-0.7}^{+0.6}$    & $9.7 $    & $9.1679_{-0.7}^{+0.6}$    & $9.5 $ \\ %% done
		       \hline
	           \fexvii\ & 2p--3d     & $15.0124_{-1.6}^{+1.9}$   & $8.6 $    & $15.0123_{-0.9}^{+0.8}$   & $7.5 $ \\%% done
               \fexvii\ & 2p--3d     & $15.2596_{-4.2}^{+1.0}$   & $4.7 $    & $15.2612_{-2.6}^{+1.9}$   & $3.7 $
  \enddata
    \tablecomments{$\lambda_{i}$ and $snr_{i}$ ($i=1$, 2) are the wavelength and signal to noise ratio of the
                   absorption lines from Methods 1 and 2. $snr_{1}=EW/\sigma_{EW}$ and $snr_{2}=a/\sigma_{a}$,
                   where $EW$ is the equivalent width and $a$ is the normalization of Gaussian-fit lines. All
                   the errors are in units of m\AA\ and at 1$\sigma$ levels.}
\end{deluxetable}
\end{tiny}
%\end{center}

\section{CORRECTION OF THE LINE CENTRAL WAVELENGTHS WITH BAYESIAN ANALYSIS}

\subsection{Bayesian Analysis}

\begin{figure}[!ht]
\center{
\includegraphics[angle=0,scale=0.50]{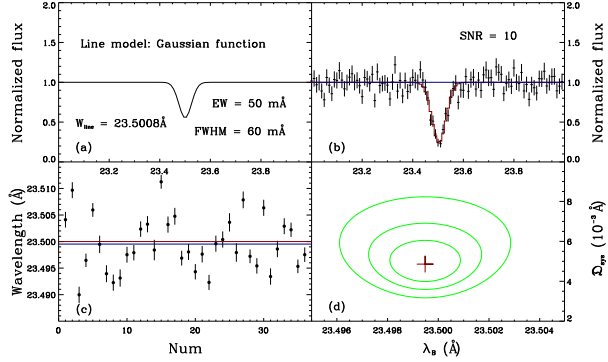}}
\caption{The simulation to illustrate how to use Bayesian analysis to obtain $\lambda_{\rm B}$ and
$\mathfrak{D}_{\rm sys}$. Panel (a) is one of the absorption line model. Panel (b) is the simulated spectrum
with the model in panel (a). The absorption line is fitted with a Gaussian function. The distribution
of 36 line wavelengths obtained from 36 simulated spectra is shown in Panel (c). The blue and red
horizontal lines are the $\overline \lambda$ and $\lambda_{\rm e}$ of $\Lambda=(\lambda_{1},~\lambda_{2},~\cdots,~\lambda_{36})$ of the 36
simulated spectra. Panel (d) presents the 2-D distribution of $\lambda_{\rm B}$ and $\mathfrak{D}_{\rm sys}$
that obtained by the Bayesian analysis to the data in Panel (c); the red cross marks the
MAP estimates of $\lambda_{\rm B}$ and $\mathfrak{D}_{\rm sys}$; the three contours from the inside
out are the 1$\sigma$, 2$\sigma$, and 3$\sigma$ credible intervals respectively, which are obtained
by calculations over a complete parameter space.}
\label{Fig:7}
\end{figure}

  For a sample consisting of $N$ observations, both Methods 1 and 2 merge all the $N$ spectra to a co-added spectrum and
can obtain the wavelength by spectral line fitting, but cannot calculate the systematic error. However, we can obtain
the line wavelengths ($\lambda_k$; $k=1$, 2, $\cdots$, $N$) and errors ($\sigma_k$; $k=1$, 2, $\cdots$, $N$) of the $N$ spectra, and then use
Bayesian analysis to obtain the 2-D probability distribution of the wavelength
($\lambda_{\rm B}$, hereafter the subscript B denotes the parameter obtained from Bayesian analysis) and the systematic
dispersion ($\mathfrak{D}_{\rm sys}$; note $\mathfrak{D}_{\rm sys}$ does not include scatter caused by statistical uncertainties).
Here we use a 2-D uniform distribution $P(\lambda_{\rm B},\mathfrak{D}_{\rm sys})$ as the {\it a prior}
distribution of $\lambda_{\rm B}$ and $\mathfrak{D}_{\rm sys}$. According to the Bayesian Theorem, the posterior
distribution $P(\lambda_{\rm B},\mathfrak{D}_{\rm sys} | \Lambda)$ is given by
\begin{equation}
P(\lambda_{\rm B},\mathfrak{D}_{\rm sys} | \Lambda)=
\frac{P(\Lambda | \lambda_{\rm B},\mathfrak{D}_{\rm sys})P(\lambda_{\rm B},\mathfrak{D}_{\rm sys})}{P(\Lambda)}
\end{equation}
where
\begin{equation}
P(\Lambda | \lambda_{\rm B},\mathfrak{D}_{\rm sys})=
\frac{1}{\sqrt{(2\pi)^{N}}}\prod_{k=1}^{N}\frac{1}{\sqrt{\mathfrak{D}_{\rm sys}^2+\sigma_{k}^2}}
e^{-\frac{(\lambda_{\rm B}-\lambda_{k})^2}{2(\mathfrak{D}_{\rm sys}^2+\sigma_{k}^2)}},
\end{equation}
\begin{equation}
P(\Lambda)=\iint P(\Lambda | \lambda_{\rm B},\mathfrak{D}_{\rm sys}) P(\lambda_{\rm B},\mathfrak{D}_{\rm sys}) d\lambda_{\rm B} d\mathfrak{D}_{\rm sys},
\end{equation}
where $\Lambda=(\lambda_1,~\lambda_2,~\cdots,~\lambda_N)$ and $\sigma_k~(k=1,~2,~\cdots,~N)$ are
the wavelength and its error of each line of these $N$ spectra.
Note that when the {\it a prior} distribution is a 2-D uniform distribution (with unspecified ranges),
the Bayesian solution is the same as that obtained from the maximum likelihood estimation.
From the 2-D probability distribution, we can obtain the maximum a posteriori (MAP) estimates and errors of $\lambda_{\rm B}$ and $\mathfrak{D}_{\rm sys}$.
$\sigma_{\rm sys}$ can be calculated by
\begin{equation}
\sigma_{\rm sys}^2=\frac{\mathfrak{D}_{\rm sys}^2}{N}.
\end{equation}

\begin{tiny}
\begin{deluxetable}{cccccc}
  %\tabletypesize{\scriptsize}
  \tablewidth{0pt}
  \tablecaption{Line parameters of the simulation in Section 4.1}
  \tablehead{  Wavelength (\AA)  &  FWHM (m\AA)  &  EW (m\AA)  &  SNR  }
  \startdata
               ($\hat{\lambda}=23.5$, $\hat{\mathfrak{D}}=0.005$)  &  60  &  50     &  10
  \enddata
  \tablecomments{The wavelength obey a Gaussian distribution ($\hat{\lambda}$, $\hat{\mathfrak{D}}$) and the other parameters are constant.}
\end{deluxetable}
\end{tiny}

As shown in Figure 7, in order to illustrate how to use Bayesian analysis to obtain $\lambda_{\rm B}$ and $\mathfrak{D}_{\rm sys}$, we make a simulation as follows:
\begin{enumerate}
\item Make 36 absorption line models (a normalized continuum plus a Gaussian absorption line) with the line parameters in Table 5.
      All the continua have spectral wavelength range ($23-24$ \AA) and resolution (0.005 \AA). One of the line model is shown in Panel (a).
\item Use the 36 models (step 1) to make 36 simulated spectra. Panel (b) is the simulated spectrum with the model in Panel (a).
\item Fit the 36 simulated spectra with a Gaussian function to obtain 36 simulated line wavelengths (Panel (c)).
\item Use Equations (7-9) to obtain the 2-D probability distribution of $\lambda_{\rm B}$ and $\mathfrak{D}_{\rm sys}$ (Panel (d)).
\end{enumerate}
From the 2-D probability distribution, we obtain the MAP estimates of $\lambda_{\rm B}$ and $\mathfrak{D}_{\rm sys}$
(red cross in Panel (c)). The maximum width of the $1\sigma$ contour in the X-axis direction is
$\sigma_{\lambda_{\rm B}}$, which is the $1\sigma$ error of $\lambda_{\rm B}$.
The expectation of the wavelength is $\lambda_{\rm e}=\hat{\lambda}\pm\hat{\mathfrak{D}}/\sqrt{N}=23.5\pm0.00083$ \AA.
The weighted average of the wavelengths ($\overline \lambda$) can be calculated by
\begin{equation}
\overline \lambda=\frac{\sum_{1}^{36}\lambda_k/\sigma_k^2}{\sum_{1}^{36}1/\sigma_k^2}.
\end{equation}
We have $\overline \lambda=23.49952\pm0.00025$ \AA\ and the results from Bayesian analysis
are $\lambda_{\rm B}=23.49948\pm0.00088$ \AA\ and $\mathfrak{D}_{\rm sys}=0.00494\pm0.00066$ \AA.
In this simple simulation, both $\overline \lambda$, $\lambda_{\rm B}$, and $\mathfrak{D}_{\rm sys}$ are consistent with the expectation.

\subsection{Comparison Between Method 1, Method 2, and Bayesian Analysis}
  In our work, three methods (Method 1, Method 2, and Bayesian analysis) can be applied to estimate the
real wavelengths of the lines from a sample. However, only the strong lines have enough observations with
enough SNRs to do the Bayesian analysis. It is very important to know which method can give  unbiased
results. We answer this question by a simulation as follows:

\begin{figure}[!ht]
\center{
\includegraphics[angle=0,scale=0.50]{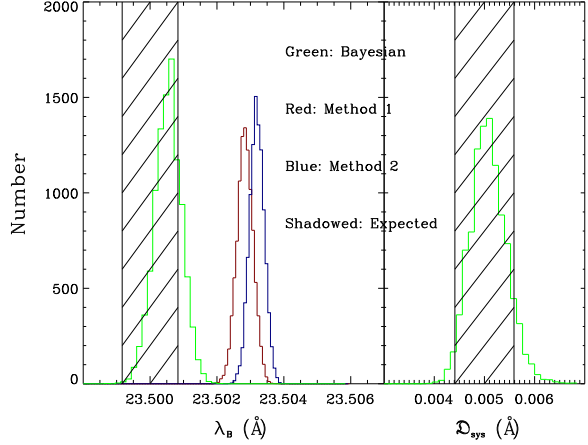}}
\caption{Comparison between the result obtained from the co-added spectra fitting (Methods 1 and 2)
and Bayesian analysis. In the left panel, the Green histogram is the distribution of $\lambda_{\rm B}$ obtained from Bayesian
analysis; the red and blue histogram are the distributions of $\lambda_{1}$ and $\lambda_{2}$ obtained from co-added
spectrum fitting (Methods 1 and 2); the shadowed region is the 1$\sigma$ confidence region of $\lambda_{\rm e}$.
In the right panel, the Green histogram is the distribution of $\mathfrak{D}_{\rm sys}$ obtained from Bayesian
analysis; the shadowed region is the 1$\sigma$ confidence region of $\mathfrak{D}_{\rm e}$.}
\label{Fig:8}
\end{figure}

\begin{tiny}
\begin{deluxetable}{cccccc}
	%\tabletypesize{\scriptsize}
	%\renewcommand\arraystretch{0.92}
  \tablewidth{0pt}
  \tablecaption{Line parameters of the simulation in Section 4.2}
  \tablehead{  $\lambda_{\rm s}$ (\AA)  &  EW (m\AA)  &  SNR  &  $\lambda_{\rm s}$ (\AA)  &  EW (m\AA)  &  SNR  }
  \startdata
                $23.5053$  &  $20.9$  &  $28.9$    &    $23.4995$  &  $14.9$  &  $11.2$    \\%% done
                $23.5091$  &  $49.9$  &  $29.7$    &    $23.4944$  &  $38.0$  &  $22.3$    \\%% done
                $23.4916$  &  $33.3$  &  $9.1 $    &    $23.4970$  &  $26.7$  &  $15.3$    \\%% done
                $23.4967$  &  $34.7$  &  $12.7$    &    $23.4926$  &  $16.6$  &  $8.7 $    \\%% done
                $23.5092$  &  $14.2$  &  $16.1$    &    $23.5016$  &  $49.3$  &  $25.3$    \\%% done
                $23.5002$  &  $23.2$  &  $14.7$    &    $23.5008$  &  $40.5$  &  $22.4$    \\%% done
                $23.4950$  &  $20.0$  &  $9.4 $    &    $23.5023$  &  $47.9$  &  $22.9$    \\%% done
                $23.4959$  &  $31.7$  &  $9.8 $    &    $23.4959$  &  $26.9$  &  $17.4$    \\%% done
                $23.4934$  &  $49.6$  &  $9.2 $    &    $23.5090$  &  $44.7$  &  $28.5$    \\%% done
                $23.4987$  &  $22.3$  &  $10.2$    &    $23.5002$  &  $28.3$  &  $8.9 $    \\%% done
                $23.5001$  &  $35.2$  &  $12.4$    &    $23.4950$  &  $29.4$  &  $18.7$    \\%% done
                $23.5037$  &  $31.4$  &  $23.0$    &    $23.5076$  &  $18.2$  &  $23.7$    \\%% done
                $23.5020$  &  $17.8$  &  $20.5$    &    $23.4957$  &  $21.8$  &  $17.1$    \\%% done
                $23.5003$  &  $22.0$  &  $26.3$    &    $23.4990$  &  $44.9$  &  $5.8 $    \\%% done
                $23.5116$  &  $33.4$  &  $25.8$    &    $23.5022$  &  $31.3$  &  $24.3$    \\%% done
                $23.5033$  &  $45.3$  &  $17.2$    &    $23.5002$  &  $27.6$  &  $20.3$    \\%% done
                $23.5044$  &  $33.4$  &  $24.3$    &    $23.4998$  &  $20.6$  &  $7.2 $    \\%% done
                $23.4961$  &  $28.3$  &  $7.7 $    &    $23.4971$  &  $40.8$  &  $18.7$
  \enddata
  \tablecomments{$\lambda_{\rm s}$ (Gaussian distribution; $\hat{\lambda}=23.5$ \AA, $\hat{\mathfrak{D}}=0.005$ \AA),
                 EWs (uniform distribution; 10-50 m\AA), and SNRs (uniform distribution; 5-30).}
\end{deluxetable}
\end{tiny}

\begin{enumerate}
\item The same as step 1 in Section 4.1, except that both EWs and SNRs
      of the 36 continua obey a uniform distribution. All the model parameters are sampled once and then fixed as shown in Table 6.
\item Repeat the steps 2--4 in Section 4.1 for 10000 times to obtain the distributions of the results from Method 1, Method 2, and Bayesian analysis (Figure 8).
\end{enumerate}
In this simulation, the wavelengths of the absorption lines ($\lambda_{s};$ $s=1,~2,~\cdots,~N$) in these 36 simulated spectra are
set to obey a Gaussian distribution ($\hat{\lambda}=23.5$ \AA, $\hat{\mathfrak{D}}=0.005$ \AA), thus the expected value of the mean value
of the wavelengths ($\lambda_{\rm e}$) should also obey a Gaussian distribution ($\hat{\lambda}$, $\hat{\mathfrak{D}}/\sqrt{N}$), i.e.,
$23.5\pm0.00083$ \AA\ (The shadowed region in the left panel of Figure 8 is the $1\sigma$ confidence region of $\lambda_{\rm e}$).
We have $\lambda_{1}=23.50284\pm0.00024$ \AA\ (Method 1; red histogram) and
$\lambda_{2}=23.50319\pm0.00022$ \AA\ (Method 2; blue histogram), which are more than $3\sigma$ deviations
from $\lambda_{\rm e}$. This is because the co-added spectra depend on the spectra with high count-rate
(Method 1) or high line-significance (Method 2), which deviate from
the expected value of 23.5 \AA\ significantly. However, the Bayesian analysis gives
$\lambda_{\rm B}=23.50054\pm0.00041$ \AA\ (green histogram) that is consistent with $\lambda_{\rm e}$.
$\mathfrak{D}_{\rm sys}$ ($0.00508\pm0.00034$ \AA) is also consistent with the expected value ($\mathfrak{D}_{\rm e}$), which obey a Gaussian distribution
($\hat{\mathfrak{D}}, \hat{\mathfrak{D}}/\sqrt{2⋅N}$), i.e., $0.005\pm0.00059$ \AA\ (The shadowed
region in the right panel of Figure 8 is the $1\sigma$ confidence region of $\mathfrak{D}_{\rm e}$).
Both the errors of $\lambda_{1}$ and $\lambda_{2}$ are smaller than that
of $\lambda_{\rm B}$; we will explain this in Section 4.3. Therefore, co-added spectral fitting can bring biased
result, but Bayesian analysis can give unbiased result for both $\lambda_{\rm B}$ and $\mathfrak{D}_{\rm sys}$.

\subsection{Correction to the Wavelength Obtained from the Co-added Spectra}
  Since the Bayesian analysis can give unbiased results to both $\lambda_{\rm B}$ and $\mathfrak{D}_{\rm sys}$,
we can use the results of the Bayesian analysis ($\lambda_{\rm B}$, $\mathfrak{D}_{\rm sys}$, and
$\sigma_{\lambda_{\rm B}}$) to correct the results obtained from the co-added spectra ($\lambda_{m}$ and
$\sigma_{\lambda_{m}}$; $m=1$, 2 for Methods 1 and 2).

  In our statistical model, the unbiased $\lambda_{\rm B}$ can be expressed as
\begin{equation}
\lambda_{\rm B}=\lambda_{m}+\Delta\lambda_{m}\pm\sigma_{\rm sys},~{m}=1,2,
\end{equation}
where $\lambda_{m}$ is the biased result, $\Delta\lambda_{m}$ is the correction quantity,
and $\sigma_{\rm sys}$ is the systematic error. Thus, $\sigma_{\lambda_{\rm B}}$ can be
written as
\begin{equation}
\sigma_{\lambda_{\rm B}}^{2}=\sigma_{\lambda_{m}}^{2}+\sigma_{\Delta\lambda_{m}}^{2}+\sigma_{\rm sys}^{2},~{m}=1,2,
\end{equation}
where $\sigma_{\lambda_{m}}$ and $\sigma_{\Delta\lambda_{m}}$ are the statistical errors
of $\lambda_{m}$ and $\Delta\lambda_{m}$ respectively. For convenience, $\sigma_{\lambda_m}$
is the mean value of the asymmetric errors of ${\lambda_m}$ in Table 4. As described in
Section 4.1, $\sigma_{\lambda_{\rm B}}$ can be obtained from the Bayesian analysis and
$\sigma_{\rm sys}$ can be calculated by Equation (10) from $\mathfrak{D}_{\rm sys}$. From Equations (12) and (13), we can obtain
\begin{equation}
\Delta\lambda_{m}=\lambda_{\rm B}-\lambda_{m},
\ {\rm and} \
\sigma_{\Delta\lambda_{m}}^{2}=\sigma_{\lambda_{\rm B}}^{2}-\sigma_{\rm sys}^{2}-\sigma_{\lambda_{m}}^{2},~{m}=1,2.
\end{equation}
From the Bayesian analysis, we can extract the statistical component ($\sigma_{\rm stat}$) from
$\sigma_{\lambda_{\rm B}}$ by
\begin{equation}
\sigma_{\rm stat}^{2}=\sigma_{\lambda_{\rm B}}^{2}-\sigma_{\rm sys}^{2},
\end{equation}
where $\sigma_{\rm stat}$ is the total statistical error that consists of both
$\sigma_{\lambda_{m}}$ and $\sigma_{\Delta\lambda_{m}}$ in our model. Therefore, we
have $\sigma_{\rm stat}>\sigma_{\lambda_{m}}$ and this is why $\sigma_{\lambda_{m}}<\sigma_{\lambda_{\rm B}}$ in Section 4.3.

  As discussed above, $\lambda_{m}$ needs to be corrected as
\begin{equation}
\lambda_{m,~\rm c}=\lambda_{m}+\Delta\lambda_{m}\pm\sigma_{\rm sys},~{m}=1,2.
\end{equation}
\begin{equation}
\sigma_{\lambda_{m,~\rm c}}^2 =\sigma_{\lambda_{m}}^2 + \sigma_{\Delta\lambda_{m}}^2 + \sigma_{\rm sys}^2,~{m}=1,2.
\end{equation}
The final errors consist of three parts: statistical error ($\sigma_{\lambda_{m}}$) from the co-added spectra,
statistical error of the correction ($\sigma_{\Delta\lambda_{m}}$), and systematic error ($\sigma_{\rm sys}$)
obtained from multi-observations. Because the systematic dispersion of the 36 spectra is partly converted into
the broadening of lines in the co-added spectrum, $\sigma_{\lambda_{m}}$ also partially includes $\sigma_{\rm sys}$.
Therefore, $\sigma_{\lambda_{m,~c}}$ is the conservative estimate of the final total error.

\begin{figure}[!ht]
\center{
\includegraphics[angle=0,scale=1.0]{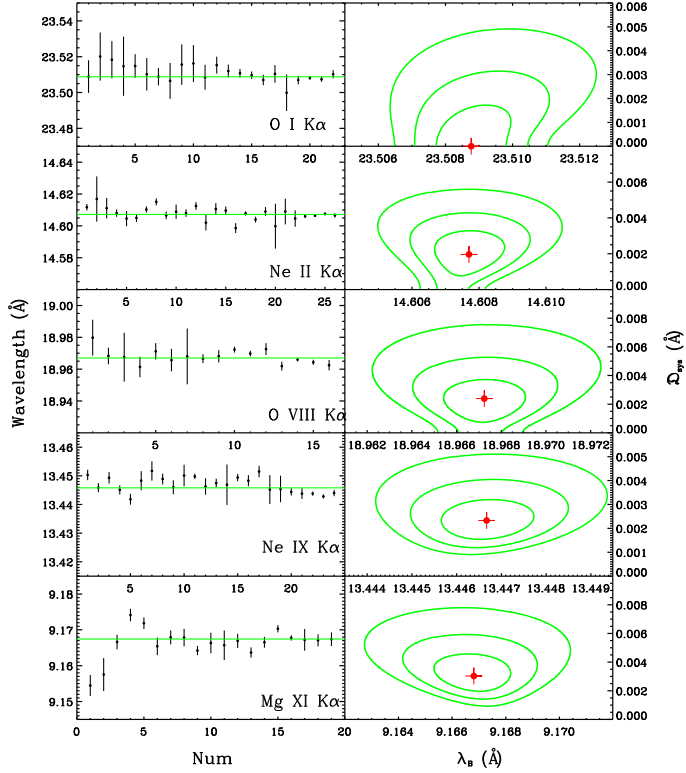}}
\caption{Distributions of the wavelengths of \oi~K$\alpha$, \neii~K$\alpha$, \oviii~K$\alpha$, \neix~K$\alpha$,
         and \mgxi~K$\alpha$ obtained from all the well-fitted observations. The contours and the red crosses
         in the right panels are the same as in Figure 7.}
\label{Fig:9}
\end{figure}

\subsection{Application of Bayesian Analysis to Our Sample}
  Although the co-added spectrum has extremely high SNR that can help us to find some weak lines,
it can bias towards the observations with high count-rate (Method 1) or high line-significance
(Method 2). Bayesian analysis can obtain the unbiased result; however, only the strong lines have
enough observations with enough SNRs to do the Bayesian analysis.

  As shown in Table 3, all the lines have been divided into five groups and the strongest line of
each group is also found. Since the absorption clouds within each group are assumed to have the
same velocities, we can obtain the unbiased result for all the lines as follows:
(1) use Methods 1 and 2 to obtain the results of high SNR co-added spectra; (2) use
the results of Bayesian analysis of
the strongest line of each group to correct all other lines in the same group, e.g., $\Delta\lambda_{m}$(\oi~K$\alpha$)
is used to correct the low-ionized O lines; see Tables 3, 7 and Figure 9 for details. The final results are shown in Tables 8--10.

The \mgxi~K$\alpha$ line is very special.
We find that for the \mgxi~K$\alpha$ line $\sigma_{\rm sys}>\sigma_{\rm stat}$. This means that, unlike \oviii~K$\alpha$ and \neix~K$\alpha$,
$\sigma_{\lambda_{1,~2}}$ of the \mgxi~K$\alpha$ lines is not dominated by statistical error, but by the components
caused by the line broadening when a co-added spectrum is produced, i.e., $\sigma_{\rm sys}$ increases
$\sigma_{\lambda_{m}}$ significantly. $\Delta\lambda_{m}$ of the \mgxi~K$\alpha$ line in Table 7 is very uncertain
and we cannot obtain $\sigma_{\Delta\lambda_{m}}$ with Equation (14). In order to correct the wavelength of the Mg lines,
we must make an assumption that the motion of both the low-ionized and the high-ionized Mg are the same as that
of Ne. Thus we can use $\Delta\lambda_{m}$(\neii~K$\alpha$) to correct the low-ionized Mg lines
and use $\Delta\lambda_{m}$(\neix~K$\alpha$) to correct the moderate-ionized and high-ionized Mg lines respectively.

For the Fe 2d-3p double lines, we also use $\Delta\lambda_{m}$(\neix~K$\alpha$) to correct the value obtained from the
co-added spectra (Methods 1 and 2; Table 4).

%\begin{center}
%\renewcommand{\arraystretch}{1.2}
\begin{tiny}
\begin{deluxetable}{lcccccccl}
      %\tabletypesize{\scriptsize}
      \tablewidth{0pt}
      \tablecaption{Comparison between the Bayesian analysis and the co-added spectrum fitting of the strong lines \oi, \oviii, \neii, \neix, and \mgxi}
      \tablehead{Ion & Transtion & $\lambda_{\rm B}$ & $\sigma_{\lambda_{\rm B}}$ & $\sigma_{\rm sys}$ & $\sigma_{\rm stat}$ &
                $\Delta\lambda_{1}$ & $\Delta\lambda_{2}$ & $\lambda_{i,~{\rm c}}$ ($i=1$, 2)}
      \startdata
           \oi\    & 1s--2p & $23.5088$ & $0.8$ & $0.0$ & $0.8$ & $+0.1 (0.5)$ & $-0.3 (0.3)$ & O (L)      \\%% done
           \oviii\ & 1s--2p & $18.9676$ & $1.1$ & $0.6$ & $1.0$ & $+0.6 (0.6)$ & $+0.8 (0.6)$ & O (M \& H) \\%% done
           \neii\  & 1s--2p & $14.6076$ & $0.7$ & $0.4$ & $0.5$ & $+0.9 (0.4)$ & $+0.6 (0.4)$ & Ne (L) \& Mg(L)     \\%% done
           \neix\  & 1s--2p & $13.4465$ & $0.7$ & $0.5$ & $0.5$ & $+1.1 (0.3)$ & $+1.3 (0.3)$ & Ne (M \& H) \& Mg (M \& H) \\%% done
           \mgxi\  & 1s--2p & $ 9.1670$ & $0.9$ & $0.7$ & $0.6$ & $-0.8 (\ast)$ & $-0.9 (\ast)$ & $\ast\ \ast\ \ast\ \ast\ $  %% done
      \enddata
      \tablecomments{$\lambda_{\rm B}$, $\sigma_{\lambda_{\rm B}}$, and $\sigma_{\rm sys}$ are obtained
                     from the Bayesian analysis in Figure 9. $\lambda_{\rm B}$ is in units of \AA\ and all
                     the other data are in units of m\AA. All the errors in parentheses are at 1$\sigma$
                     levels. $\sigma_{\rm stat}$ is calculated with Equation (15). $\Delta\lambda_{1}$ and $\Delta\lambda_{2}$,
                     as well as their errors, are calculated with Equation (14). `$\ast$' means that we cannot obtain
                     $\sigma_{\Delta\lambda_{i}}$ because $\sigma_{\rm stat}$ is smaller than $\sigma_{i}$
                     in Table 4. The last column lists the lines to be corrected and `L', `M' and `H'
                     are the same as that defined in Table 3. Please see Section 4.3 for details.}
\end{deluxetable}
\end{tiny}
%\end{center}

%\begin{center}
%\renewcommand{\arraystretch}{1.2}
\begin{small}
\begin{deluxetable}{lccccccc}
  %\tabletypesize{\scriptsize}
  \tablewidth{0pt}
  \tablecaption{Comparison of the wavelengths of low-ionized elements
                between the corrected values in this paper, observations in Y09, and the theoretical calculations.}
  \tablehead{  Ion      & Transtion  & $\lambda_{\rm 1,~c}$(\AA)  & $\lambda_{\rm 2,~c}$(\AA) & $\lambda_{\rm m}$(\AA) & $\rm Y09$(\AA) & $\rm G05BN02$(\AA) & $\rm G00GM05$(\AA)}
  \startdata
               \oi\     & 1s--2p     & $23.5088_{-0.8}^{+0.8}$   & $23.5088_{-0.8}^{+0.8}$  & $23.5088_{-0.8}^{+0.8}$  & $23.508_{-1.6}^{+1.6}$  & $23.4475$      & $23.532$   \\%% done
               \oi\     & 1s--3p     & $22.8834_{-2.4}^{+2.4}$   & $22.8869_{-2.3}^{+2.4}$  & $22.8852_{-2.4}^{+2.4}$  & $22.886(f)$             & $\cdots$       & $22.907$   \\%% done
               \oii\    & 1s--2p     & $23.3508_{-1.0}^{+1.4}$   & $23.3503_{-1.1}^{+1.2}$  & $23.3505_{-1.1}^{+1.3}$  & $23.348_{-4.2}^{+4.2}$  & $23.310$       & $\cdots$   \\%% done
               \oiii\   & 1s--2p     & $23.0392_{-5.4}^{+8.5}$   & $23.0562_{-8.3}^{+7.1}$  & $23.0477_{-7.0}^{+7.8}$  & $\cdots$                & $23.0692^{a}$  & $\cdots$   \\%% done
               \hline
               \nei\    & 1s--3p     & $14.2946_{-3.0}^{+1.1}$   & $14.2948_{-0.9}^{+0.9}$  & $14.2947_{-2.2}^{+1.0}$  & $14.294_{-1.3}^{+1.5}$  & $\cdots$       & $14.298$   \\%% done
               \neii\   & 1s--2p     & $14.6077_{-0.7}^{+0.7}$   & $14.6077_{-0.7}^{+0.7}$  & $14.6077_{-0.7}^{+0.7}$  & $14.605_{-1.0}^{+1.0}$  & $14.631$       & $14.605$   \\%% done
               \neii\   & 1s--3p     & $14.0038_{-0.9}^{+1.2}$   & $14.0037_{-1.1}^{+1.1}$  & $14.0037_{-1.0}^{+1.2}$  & $14.001_{-1.2}^{+2.0}$  & $14.0069^{a}$  & $\cdots$   \\%% done
               \neii\   & 1s--4p     & $13.9382_{-3.1}^{+1.6}$   & $13.9345_{-3.0}^{+3.3}$  & $13.9363_{-3.1}^{+2.6}$  & $\cdots$                & $13.9393^{a}$  & $\cdots$   \\%% done
               \neiii\  & 1s--2p     & $14.5077_{-0.8}^{+0.9}$   & $14.5079_{-0.9}^{+0.9}$  & $14.5078_{-0.9}^{+0.9}$  & $14.507_{-2.1}^{+2.0}$  & $14.526$       & $14.518$   \\%% done
               \neiii\  & 1s--3p     & $13.6960_{-2.4}^{+1.5}$   & $13.6959_{-2.5}^{+1.8}$  & $13.6959_{-2.5}^{+1.7}$  & $13.690_{-1.5}^{+6.3}$  & $13.6977^{a}$  & $\cdots$   \\%% done
               \hline
               \mgiii\  & 1s--3p     & $9.4339_{-22.1}^{+44.8}$  & $9.4772_{-4.2}^{+1.0}$   & $9.4555_{-15.9}^{+31.7}$ & $\cdots$                & $9.4504^{a}$   & $\cdots$
  \enddata
    \tablecomments{$\lambda_{i,~{\rm c}}$ ($i=1$, 2) are the corrected value of the low-ionized elements in
                   Table 4. $\lambda_{\rm m}$ is the average value of $\lambda_{i,~c}$.
                   The O lines are corrected by $\Delta\lambda_{i}$ of \oi~K$\alpha$ in Table 7,
                   and the Ne and Mg lines are both corrected by that of \neii~K$\alpha$.
                   $\sigma_{\lambda_{i,~c}}$ are asymmetric; the upper and lower 1$\sigma$ errors are calculated by
                   $\sigma_{\lambda_{i,~c},~{\rm j}}^2=\sigma_{\lambda_{i},~{\rm j}}^2+\sigma_{\Delta\lambda_{i}}^2+\sigma_{\rm sys}^2$ ($j=$u, l).
                   Values of oxygen in columns
                   7 and 8 are from Garc\'ia et al. (2005) and Gorczyca (2000) respectively. Values of neon in columns 7 and 8
                   are from Behar \& Netzer (2002) and Gorczyca \& McLaughlin (2005) respectively. All the errors are in units
                   of m\AA\ at 1$\sigma$ level and `$f$' in parentheses means `fixed'. \\
                   $^a$ The value obtained from M. Gu (2010, private communication; hereafter Gu10)}
\end{deluxetable}
\end{small}
%\end{center}

%\begin{center}
%\renewcommand{\arraystretch}{1.2}
\begin{tiny}
\begin{deluxetable}{lccccccc}
  %\tabletypesize{\scriptsize}
  \tablewidth{0pt}
  \tablecaption{Comparison of the wavelengths of moderate-ionized elements
                between the corrected values in this paper, observations in Y09, and the theoretical calculations.}
  \tablehead{  Ion      & Transtion  & $\lambda_{\rm 1,~c}$(\AA)  & $\lambda_{\rm 2,~c}$(\AA) & $\lambda_{\rm m}$(\AA) & $\rm Y09$(\AA) & $\rm Gu10$(\AA)}
    \startdata
               \oiv\    & 1s--2p     & $22.6975_{-37.3}^{+3.1}$  & $22.6964_{-3.4}^{+3.4}$   & $22.6969_{-26.5}^{+3.3}$  & $\cdots$                & $22.7515$  \\%% done
               \ov\     & 1s--2p     & $22.2855_{-3.4}^{+4.8}$   & $22.2868_{-4.2}^{+3.0}$   & $22.2861_{-3.8}^{+4.0}$   & $\cdots$                & $22.3682$  \\%% done
               \ovi\    & 1s--2p     & $22.0287_{-1.7}^{+2.6}$   & $22.0289_{-2.2}^{+1.9}$   & $22.0288_{-2.0}^{+2.3}$   & $22.026_{-4.0}^{+4.0}$  & $22.0403$  \\%% done
               \hline
               \neiv\   & 1s--2p     & $14.3482_{-64.3}^{+64.3}$ & $\cdots$                  & $14.3482_{-64.3}^{+64.3}$ & $\cdots$                & $14.3710$   \\%% done
               \nev\    & 1s--2p     & $14.2138_{-38.6}^{+10.2}$ & $14.2096_{-13.7}^{+12.6}$ & $14.2117_{-29.0}^{+11.5}$ & $\cdots$                & $14.2126$   \\%% done
               \nevii\  & 1s--2p     & $13.8284_{-2.0}^{+4.2}$   & $13.8275_{-4.3}^{+3.3}$   & $13.8279_{-3.4}^{+3.8}$   & $\cdots$                & $13.8262$   \\%% done
               \hline
               \mgiv\   & 1s--2p     & $9.8550_{-1.3}^{+29.1}$   & $\cdots$                  & $9.8550_{-1.3}^{+29.1}$   & $\cdots$                & $9.8786$   \\ %% done
               \mgv\    & 1s--2p     & $9.8202_{-30.1}^{+30.1}$  & $9.8210_{-2.1}^{+1.6}$    & $9.8206_{-21.3}^{+21.3}$  & $\cdots$                & $9.8034$   \\ %% done
               \mgv\    & 1s--3p     & $9.0581_{-2.1}^{+28.0}$   & $9.0559_{-2.1}^{+2.6}$    & $9.0570_{-2.1}^{+19.9}$   & $\cdots$                & $9.0766$   \\ %% done
               \hline
               \fexvii\ & 2p--3d     & $15.0135_{-1.8}^{+2.0}$   & $15.0130_{-1.1}^{+1.0}$   & $15.0133_{-1.5}^{+1.6}$   & $15.010(f)$                & $15.015^{a}$   \\ %% done
               \fexvii\ & 2p--3d     & $15.2607_{-4.3}^{+1.2}$   & $15.2620_{-4.6}^{+2.0}$   & $15.2614_{-4.5}^{+1.6}$   & $\cdots$                & $15.262^{a}$
      \enddata
    \tablecomments{The same as Table 8, but $\lambda_{i,~{\rm c}}$ ($i=1$, 2) is the corrected wavelength of the moderate-ionized elements in Table 4.
                   The O lines are corrected by $\Delta\lambda_{i}$ of \oviii~K$\alpha$ in Table 7,
                   and the Ne, Mg, and Fe lines are both corrected by that of \neix~K$\alpha$.
                   The theoretical calculations of O, Ne, and Mg are obtained from Gu10. \\
                   $^a$ The value obtained from NIST.}
\end{deluxetable}
\end{tiny}
%\end{center}

%\begin{center}
%\renewcommand{\arraystretch}{1.2}
\begin{tiny}
\begin{deluxetable}{lccccccc}
  %\tabletypesize{\scriptsize}
  \tablewidth{0pt}
  \tablecaption{Comparison of the wavelengths of high-ionized elements
                between the corrected values in this paper, observations in Y09, and the theoretical calculations.}
  \tablehead{  Ion      & Transtion  & $\lambda_{\rm 1,~c}$(\AA)  & $\lambda_{\rm 2,~c}$(\AA) & $\lambda_{\rm m}$(\AA) & $\rm Y09$(\AA) & $\rm NIST$(\AA) & $\rm V96$(\AA)}
    \startdata
               \ovii\   & 1s--2p     & $21.5921_{-1.5}^{+1.8}$   & $21.5957_{-1.8}^{+1.7}$   & $21.5939_{-1.7}^{+1.8}$   & $21.602(f)$              & $21.6020$     & $21.6019$   \\%% done
               \ovii\   & 1s--3p     & $18.6265_{-1.1}^{+1.4}$   & $18.6263_{-1.3}^{+1.4}$   & $18.6264_{-1.2}^{+1.4}$   & $18.625_{-2.5}^{+2.6}$   & $18.6270$     & $18.6288$   \\%% done
               \ovii\   & 1s--4p     & $17.7663_{-1.5}^{+1.3}$   & $17.7682_{-1.5}^{+1.2}$   & $17.7673_{-1.5}^{+1.3}$   & $17.765(f)$              & $\cdots$      & $17.7686$   \\%% done
               \oviii\  & 1s--2p     & $18.9673_{-1.3}^{+1.0}$   & $18.9673_{-1.1}^{+1.1}$   & $18.9673_{-1.2}^{+1.1}$   & $18.964_{-1.7}^{+2.0}$   & $18.9689^{\ast}$ & $18.9689^{\ast}$ \\%% done
               \oviii\  & 1s--3p     & $16.0052_{-1.6}^{+1.1}$   & $16.0053_{-1.5}^{+1.6}$   & $16.0052_{-1.6}^{+1.4}$   & $16.003_{-6.7}^{+6.7}$   & $16.0059^{\ast}$ & $16.0059^{\ast}$ \\%% done
               \hline
               \neix\   & 1s--2p     & $13.4466_{-0.7}^{+0.7}$   & $13.4466_{-0.7}^{+0.7}$   & $13.4466_{-0.7}^{+0.7}$   & $13.445_{-1.2}^{+1.1}$   & $13.4470$     & $13.4471$   \\%% done
               \neix\   & 1s--3p     & $11.5470_{-0.9}^{+1.5}$   & $11.5469_{-1.1}^{+1.7}$   & $11.5469_{-1.0}^{+1.6}$   & $11.549_{-3.4}^{+1.4}$   & $11.5470$     & $11.5466$   \\%% done
               \neix\   & 1s--4p     & $11.0021_{-2.6}^{+2.6}$   & $11.0000_{-1.9}^{+2.0}$   & $11.0010_{-2.3}^{+2.3}$   & $\cdots$                 & $11.0128^{a}$ & $\cdots$   \\%% done
               \nex\    & 1s--2p     & $12.1261_{-4.1}^{+2.2}$   & $12.1266_{-2.5}^{+3.0}$   & $12.1264_{-3.4}^{+2.6}$   & $12.134(f)$              & $\cdots$      & $12.1339^{\ast}$ \\%% done
               \hline
               \mgxi\   & 1s--2p     & $9.1690_{-1.0}^{+1.0}$    & $9.1685_{-1.1}^{+1.0}$    & $9.1687_{-1.1}^{+1.0}$    & $9.170_{-1.2}^{+0.6}$    & $9.1689$      & $9.1688$
      \enddata
    \tablecomments{The same as Table 8, but $\lambda_{i,~{\rm c}}$ ($i=1$, 2) is the corrected wavelength of the high-ionized
                   elements in Table 4. The O lines are corrected by $\Delta\lambda_{i}$ of \oviii~K$\alpha$ in Table 7, and
                   the Ne and Mg lines are both corrected by that of \neix~K$\alpha$.\\
                   $^a$ The value obtained from Gu10.\\
                   $^{\ast}$ The values of the wavelengths of these lines are the weighted centroids of the
                   doublets in the databases listed in this table.}
\end{deluxetable}
\end{tiny}
%\end{center}

\section{DISCUSSION}
\subsection{The Two Methods of Merging the Spectra}
  In the Section 3, two different methods are used to merge all the 36 observations. Method 1 is very simple:
we only need to add the counts of each spectrum channel by channel, i.e., every spectrum has the same weight.
However, Method 1 implicitly assumes that all the spectra have the same absorption column densities, i.e., the
same weight. When we merge two spectra that one has a strong absorption line (e.g., the \ovi~K$\alpha$ in ObsId
6618) and the other has a weak absorption line (e.g., the \ovi~K$\alpha$ in ObsId 8599), the spectrum with weak
absorption line can dilute or even wipe out the absorption line in the co-added spectrum, i.e., reduce the
significance of the absorption line. To avoid this problem, we use the significance of the absorption line as the
weight to merge the spectra instead (Method 2). The spectra that only contribute to the continuum are abandoned
in this method, since the continuum is actually the noise for the absorption lines. From Table 4, we can see that
the results obtained by the two methods are consistent with each other for most of the absorption lines, except
for some weak lines (e.g., \oiii~K$\alpha$) that are too weak to be fit with a Gaussian function. In addition,
several suspicious lines may be faked, as they are only present in the co-added spectrum of Method 1 (\mgiv~K$\alpha$
and \neiv~K$\alpha$). The significance levels of the lines in the co-added spectrum obtained with
Method 2 are not remarkably higher than that obtained with Method 1. This means that the assumptions of both
Methods 1 and 2 are reasonable for our sample. More high-quality observations of more targets are the key to the
determination of the wavelength of these weak transitions.

\subsection{Systematic Errors of the Lines in Co-added Spectra}
  As described in Section 4.2, although both Methods 1 and 2 can obtain the spectra with high SNRs, however,
both $\lambda_{1}$ and $\lambda_{2}$ are biased. In addition, $\sigma_{\lambda_{1}}$ and $\sigma_{\lambda_{2}}$
are smaller than the true value, because $\sigma_{\rm sys}$ is not considered.
In this sub-section, we emphasize another important source of $\sigma_{\rm sys}$, i.e.,
$\sigma_{\rm sys}$ caused by the uncertainty of the spectral fitting.

\begin{figure}[!ht]
\center{
\includegraphics[angle=0,scale=0.60]{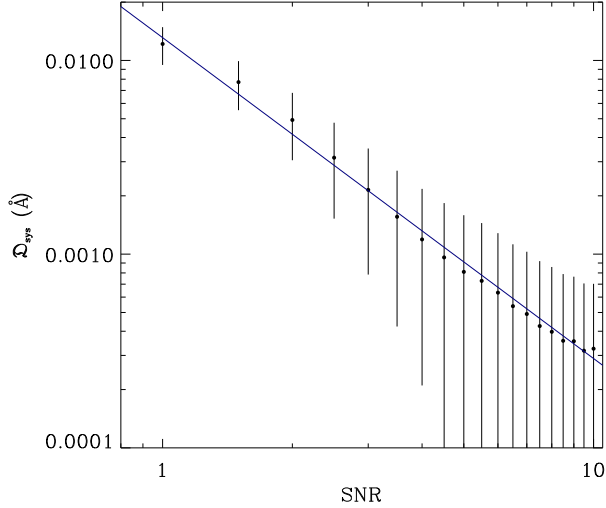}}
\caption{The dependence of $\mathfrak{D}_{\rm sys}$ on the SNRs of the spectra.}
\label{Fig:10}
\end{figure}

  The SNRs of some observations are so low that the fluctuations of the continua will seriously
affect the fitting result of the weak lines. We can quantity this uncertainty with simulation. In this
simulation, we take the SNR of the simulated spectrum as the only variable to test the dependence of
the systematic uncertainty on the SNRs of the spectra. The SNR range is between 1 and 10, with a step
of 0.5, as shown in Figure 10. For each SNR, we make the simulation as follows:
\begin{enumerate}
\item The same as step 1 in Section 4.1, except that the line central wavelength is fixed at 23.5 \AA.
\item Repeat the steps 2--4 in Section 4.1 10000 times to obtain $\lambda_{\rm B}$ and the error of the $\mathfrak{D}_{\rm sys}$.
\end{enumerate}
Finally, for each SNR, we obtain $\mathfrak{D}_{\rm sys}$ and its error, as shown in Figure 10. We
find that $\mathfrak{D}_{\rm sys}$ increases dramatically as the SNR decreases.
The relationship between the SNR and $\mathfrak{D}_{\rm sys}$ is similar to the form of a power-law.
The reason for this anti-correlation is that the error propagation of the Gaussian fitting is nonlinear
when the statistical error is large. Therefore, the joint analysis of all the observations can reduce
the systematic uncertainty due to the improvement of the SNRs of the co-added spectra.

$\sigma_{\rm sys}$ can be caused by other uncertainties, such as the imprecise Galactic rotation
correction that depends on several uncertain models (e.g., the distribution of the gas inside and above
the Galactic Plane). Because the parameters of the Galactic rotation model are still uncertain, we cannot give accurate value
of systematic uncertainty caused by this uncertainty. Nevertheless, the total systematic uncertainty can
be obtained from the distribution of the lines of the 36 observations, as described in Section 4.1. Due to the
limitation of the SNRs, we cannot do Bayesian analysis to all the lines. However, since the absorption
clouds with the similar degree of ionization are assumed to have the same velocities
as shown in Table 3, we use $\sigma_{\rm sys}$ of the strongest line to represent the systematic uncertainty of
all the lines in the same group, e.g., $\sigma_{\rm sys}$(\neii~K$\alpha$) represents the systematic uncertainty
of the low-ionized Ne lines.

\begin{figure}[!ht]
\center{
\includegraphics[angle=0,scale=1.0]{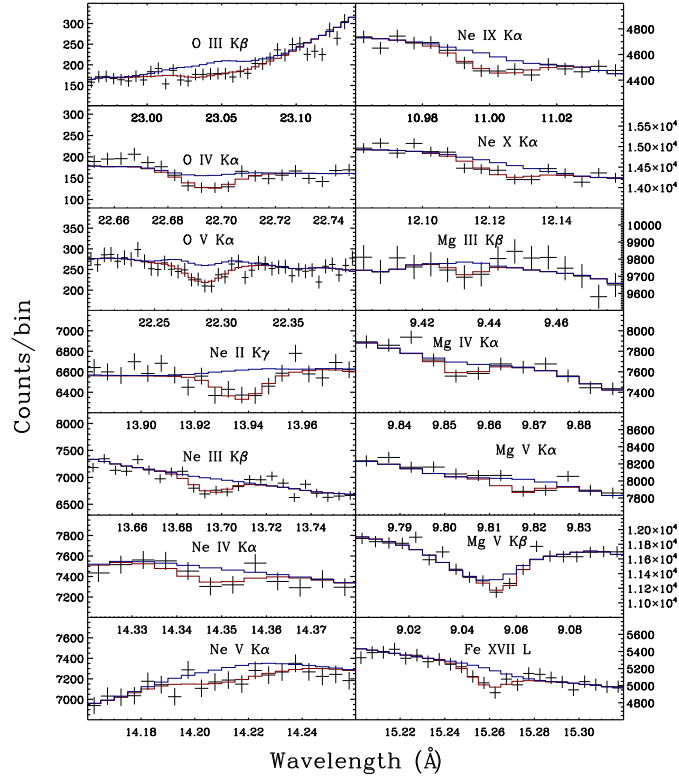}}
\caption{All the lines with ${snr_1}<5$ in Table 4.}
\label{Fig:11}
\end{figure}

\subsection{Detection of the K Transitions in Soft X-ray Band}
  As described above, our results on the wavelengths of K transitions of high-ionized O, Ne, and Mg are
the most accurate so far, because more sources are used to reduce the systematic uncertainty induced by
the Galactic rotation correction and more observations are used to reduce the statistical error. Our
results are also unbiased, because we do the bias correction for each line. Among all the lines,
\neix~K$\alpha$ is the most accurate, with an error of only 0.7 m\AA, which is equivalent to about 14
$\rm km~s^{-1}$. This error is so small that it may be used to measure the low-velocity gas, e.g., the hot
high-ionized gas in Galactic halo. For the lines whose wavelengths are consistent between NIST and V96,
our results are also consistent with those. For those lines whose wavelengths are inconsistent between
NIST and V96, our results are closer to that in NIST. Compared to Y09, our results are two times more
accurate (e.g., \oviii~K$\alpha$) and are also consistent with theoretical calculations except the \ovii~K$\alpha$
and the \nex~K$\alpha$ lines (Table 10). In our work, $\lambda_{\rm m}$(\nex~K$\alpha)=12.1264_{-3.4}^{+2.6}$ \AA\
is 7.5 m\AA\ lower than the value of 12.1339 \AA\ in V96. We note that NIST does not include this line.
The \ovii~K$\alpha$ line is more complicated and it is also 10 m\AA\ lower than that in NIST and V96.
We will discuss this in the next sub-section.

  Unlike the high-ionized elements, in the soft X-ray band, there still exists large discrepancies between
the astronomical observations and the theoretical calculations of K transitions of the low-ionized and moderate-ionized
elements (Table 1). In fact, some of these lines (e.g., 1s--3p of \mgiii\ and 1s--2p of \mgiv) have even never been detected.
The co-added spectra have excellent SNRs that provides us an unique opportunity to detect these weak absorption lines.
We are particularly concerned about these transitions and thus search for the signals of these absorption lines around
their theoretical values. For the 11 transitions of the low-ionized elements shown in Table 8, $N_{snr_1>5}=7$ and
$N_{3<snr_1<5}=3$. Only ${snr_1}$(\mgiii~K$\alpha)<3$, but ${snr_2}$(\mgiii~K$\alpha)>3$. Despite the low SNRs, it is
the first detection of the \mgiii~K$\alpha$ line so far. Several moderate-ionized ions, i.e., the O, Ne, and Mg lines shown
in Table 9 (except the \ovi~K$\alpha$ that has been detected in Y09), also appear for the first time around the theoretical
values in the co-added spectra. As shown in Table 9, only ${snr_1}$(\ovi~K$\alpha$ and \nevii~K$\alpha)>5$. However, among
the other lines, ${snr_1}$(\neiv~K$\alpha$ and \mgiv~K$\alpha)\leq3$. Moreover, the \neiv~K$\alpha$ and \mgiv~K$\alpha$
lines are only present in the co-added spectrum of Method 1, which hints that these three
lines may be artificial and need to be confirmed by higher quality observations in the future.

In Y09, only one of the 2d-3p double lines, i.e., the \fexvii\ 15.01 \AA\ line was detected, but no statistical
error was given. In our work, both \fexvii\ 15.01 \AA\ and 15.26 \AA\ lines are detected significantly for the
first time simultaneously (Figures 3 and 4; Table 9). All the lines with ${snr_1}<5$ are shown in Figure 11.

\begin{figure}[!b]
\center{
\includegraphics[angle=0,scale=0.60]{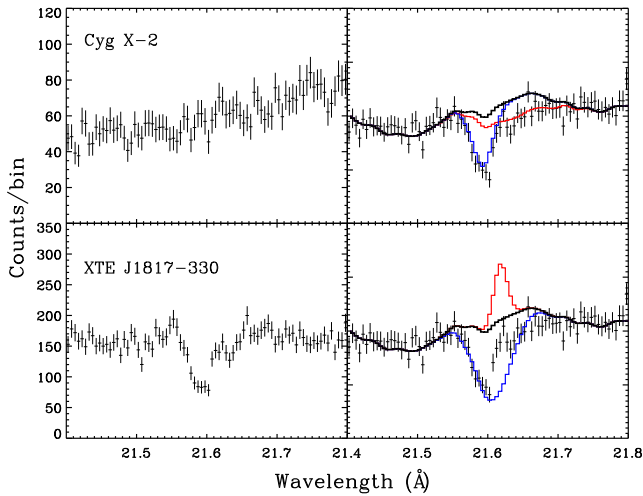}}
\caption{The left two panels are the co-added spectra around \ovii~K$\alpha$ of Cyg X--2 and
         XTE J1817--300 respectively. The line fitting of the final co-added spectrum with two
         different models (upper: absorption + absorption; lower: absorption + emission) are
         shown in the right panels.}
\label{Fig:12}
\end{figure}

\subsection{Wavelength of the {\textbf \rm \ovii~K$\alpha$} Lines}
  By analyzing  the co-add spectrum, the wavelengths of all the strong absorption lines
can be determined accurately. In our work, several \ovii\ lines, including even the very
weak \ovii~K$\gamma$, are found. However, the wavelength of \ovii~K$\alpha$ is still
uncertain. The observations of Cyg X--2 and XTE J1817--330 are the most important components
of the co-added spectrum, which have enough counts and thus can greatly determine the shape
of the co-added spectrum. Y09 jointly analyzed four {\sl Chandra}-HETG observations of Cyg X--2 and
found the \ovii~K$\beta$ line clearly but no \ovii~K$\alpha$ line was detected. This problem
can be interpreted by an unknown emission that fills in the absorption (Cabot et al. 2013).
The spectrum of  XTE J1817--330 is similar to Cyg X--2 but more complex, which can be caused
by not only two absorption lines but also an absorption line plus an emission line. Thus, the
co-added spectrum can also be well fitted by two different models (Figure 12). For the model
of a strong absorption line plus a weak redshifted absorption line, the fitting value of the
\ovii~K$\alpha$ line is $21.5915\pm0.0015$ \AA. However, for the model of an absorption line
plus a weak emission line, the fitting value is $21.6074\pm0.0015$ \AA. Both values are very
different from the theoretical value of 21.6020 \AA\ given by NIST or 21.6019 \AA\ given by
V96. We cannot distinguish which model is correct due to the limitation of the SNR of the
spectrum.

\section{SUMMARY}
  So far, the wavelengths of K alpha absorption lines of neutral, low-ionized, and moderate-ionized O, Ne, and Mg
have not been determined precisely, either in theoretical calculations (e.g., V96; NIST),
laboratory measurements (Stolte et al. 1997), or in astronomical observations (J0406; Y09).
In order to obtain the wavelengths of K transitions of O, Ne, and Mg more accurately, we
jointly analyzed 36 {\sl Chandra}-HETG observations of 11 LMXBs at low Galactic latitudes in this
work. We corrected the Galactic rotation velocity to the rest frame for every observation and
then used two different methods to merge all the corrected spectra to a co-added spectrum.
Finally, we fit and obtained the wavelengths of every absorption line in the co-added spectrum
obtained by the above two methods (Section 3). Both methods give similar and consistent results
for most lines, as shown in Figure 3.

  We noted that the co-added spectra obtained by the usual methods (Methods 1 and 2) exhibit
biases, which are very important for the calibration of the lines. The co-added spectrum can
be biased to the observations of high counts (Method 1) or high SNRs of the lines (Method 2).
We made Bayesian analysis to the lines of the \oi~K$\alpha$, \neii~K$\alpha$, \oviii~K$\alpha$,
\neix~K$\alpha$ to obtain the systematic uncertainty and the bias correction of the O, Ne, and
Mg lines in all phases. The final results after the bias correction are summarized as follows:

(1) For the the neutral, low-ionized, and high-ionized lines, the accuracy of our result is
five and two times higher than J0406 and Y09 respectively (Table 4). Several lines that were
not detected (i.e.,\oiii~K$\alpha$, \neii~K$\gamma$, \mgiii~K$\beta$, and \neix~K$\gamma$) or too
weak to give measurement errors (i.e., \oi~K$\beta$, \ovii~K$\gamma$ and \nex~K$\alpha$) in Y09
are detected clearly in our work. We also find the moderate-ionized lines of these elements
(\oiv~K$\alpha$, \ov~K$\alpha$, \neiv~K$\alpha$, \nev~K$\alpha$, \nevii~K$\alpha$, \mgiv~K$\alpha$, \mgv~K$\alpha$, and \mgv~K$\beta$;
Table 9) whose significances are so low that need to be confirmed in the future. Besides the
remarkable improvement of the accuracy, it is worth mentioning that all the biases measurements
here are corrected (Tables 8--10).

(2) The systematic uncertainty in our measurement mainly comes from the Galactic rotation correction
and the spectral fitting. The former depends on the Galactic model. For the latter, we make simulations
to estimate the effect. The total systematic uncertainties are: $\sigma_{\rm sys}=0$ m\AA\ and 0.6 m\AA\
for low-ionized and high-ionized O, $\sigma_{\rm sys}=$ 0.4 m\AA\ and 0.5 m\AA\ for low-ionized and high-ionized Ne,
$\sigma_{\rm sys}=0.7$ m\AA\ for Mg lines (Table 7).

(3) For high-ionized lines of \ovii~K$\beta$ and \oviii~K$\alpha$, our results are consistent with
that of NIST, but 2 m\AA\ lower than V96. For \nex~K$\alpha$, our result is about 8 m\AA\ lower than
that given by V96; NIST does not provide the theoretical value. For the moderate-ionized lines, the
discrepancy between our measurements and the theoretical calculations are generally between 1 to
80 m\AA. Because the statistical errors of these lines are also similar to the discrepancy,
the measurements are consistent with the theoretical calculations.

\acknowledgments JYL thanks Dr. Yuan Liu for helping on improving the draft manuscript. SNZ acknowledges partial
funding support by 973 Program of China under grant 2009CB824800, and by the National Natural Science Foundation
of China under grant Nos. 11133002, 10821061, and 10725313.

\vspace{36pt}

\appendix

\centerline {\textbf {\textbf APPENDICES}}

\section{The relationship between the absorption column density ($N_{\rm H}$) and the error of the wavelength of
the absorption line ($\sigma_{\lambda}$)}
\label{A}

For the spectrum with a continuum plus a Gaussian absorption line, the wavelength of the
line center ($\lambda$) can be determined by line fitting with a Gaussian function
\begin{equation}
\phi(\lambda_i) = d e^{-\frac{(\lambda_i-\lambda)^2}{2b^2}},
\end{equation}
where $d$, $b$, and $\lambda$ are the depth, broadening, and center wavelength of the line.
As described in Landman et al. (1982) and Lenz \& Ayres (1992), $\sigma_{\lambda}$ (the error of $\lambda$)
can be expressed as
\begin{equation}
\sigma_{\lambda}^2 \propto \frac{\sigma^2 b}{d^2},
\end{equation}
where $\sigma$ is the error of the continuum around the line.
For an observed spectrum, if the observation time is fixed, $\sigma^2$ and $d$ can be expressed as
\begin{equation}
\sigma^2 \propto f \propto e^{-\tau_{\rm c}}, \tau_{\rm c} = \sigma_{\rm c} N_{\rm H},
\end{equation}
and
\begin{equation}
d \propto f \cdot (1-e^{-\tau_{\rm l}}) \propto e^{-\tau_{\rm c}} (1-e^{-\tau_{\rm l}}), \tau_{\rm l} = \sigma_{\rm l} f_{\rm i} N_{\rm H},
\end{equation}
where $f$ is the flux of the continuum, $\tau_{\rm c}$ and $\tau_{\rm l}$ are the optical depths of the continuum and the line center,
$\sigma_{\rm c}$ and $\sigma_{\rm l}$ are the cross sections for the continuum and the line center,
and $f_{\rm i}$ is the abundance of the ion producing the absorption line.
Thus $\sigma_{\lambda}^2$ can be written as
\begin{equation}
\sigma_{\lambda}^2 \propto \frac{1}{e^{-\sigma_c N_{\rm H}}(1-e^{-\sigma_{\rm l} f_{\rm i} N_{\rm H}})^2}.
\end{equation}

We use the observation of the \oviii~K$\alpha$ line in Y09 to show the relationship between
$N_{\rm H}$ and $\sigma_{\lambda}$. Here $\sigma_{\rm c} = 5.9 \times 10^{-22} {\rm cm}^2$ is obtained
from Morrison \& McCammon (1983) and $\sigma_{\rm l} f_{\rm i} = 1.1 \times 10^{-22} {\rm cm}^2$ is calculated
from the observation in Y09. As shown in Figure 13, for either too low or too high $N_{\rm H}$ the error of the
\oviii~K$\alpha$ can be increased significantly.

\begin{figure}[!t]
\center{
\includegraphics[angle=0,scale=0.6]{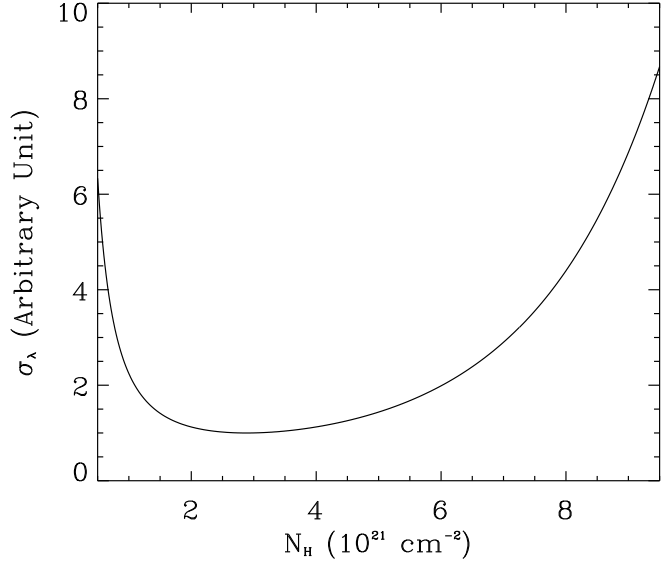}}
\caption{The relationship between the absorption column density ($N_{\rm H}$) and
         the error of the wavelength of the absorption line ($\sigma_{\lambda}$).
         The lowest point of the curve is normalized to 1.}
\label{Fig:13}
\end{figure}

\section{The weights of the absorption lines to merge to a co-added spectrum.}
\label{B}

In Equation (A2), $d$ can also be described by $d \propto \frac{a}{b}$, where $a$
is the normalization of the line. Thus we have
\begin{equation}
\sigma_{\lambda}^2 \propto \frac{\sigma^2 b^3}{a^2}.
\end{equation}
When merging two spectra with the same absorption lines with weights $w_1$ and $w_2$, we have
\begin{equation}
\sigma'^2=\sigma_1^2 w_1^2 + \sigma_2^2 w_2^2
\end{equation}
\begin{equation}
a' = a_1 w_1 + a_2 w_2
\end{equation}
\begin{equation}
d' = d_1 w_1 + d_2 w_2 \propto \frac{a_1}{b_1}w_1 + \frac{a_2}{b_2}w_2
\end{equation}
\begin{equation}
b' \propto \frac{a'}{d'} \propto \frac{a_1 w_1+a_2 w_2}{\frac{a_1}{b_1} w_1 + \frac{a_2}{b_2} w_2}.
\end{equation}
where superscript `$'$' refers to the parameters of the co-add spectra. Thus ${\sigma'_\lambda}^2$ can be expressed as
\begin{equation}
{\sigma'_\lambda}^2 \propto \frac{\sigma'^2 b'^3}{a'^2} \propto (\sigma_1^2 w_1^2 + \sigma_2^2 w_2^2) \frac{a_1 w_1 + a_2 w_2}{(\frac{a_1}{b_1} w_1 + \frac{a_2}{b_2} w_2)^3}.
\end{equation}
Defining $k=\frac{w_1}{w_2}$, $m=\frac{a_1}{a_2}$, and $n=\frac{b_1}{b_2}$, we then have
\begin{equation}
{\sigma'_\lambda}^2 \propto (\sigma_1^2 k^2 + \sigma_2^2) \frac{m k + 1}{(\frac{m}{n}k + 1)^3}.
\end{equation}
Usually the absorption lines to be merged usually have a similar broadening, i.e., $n \sim 1$,
as exemplified in Figure 14. Therefore, the above equation can be simplified as
\begin{equation}
{\sigma'_\lambda}^2 \propto (\sigma_1^2 k^2 + \sigma_2^2) \frac{1}{(\frac{m}{n}k + 1)^2}.
\end{equation}
We can obtain that the value of $k$ for minimum ${\sigma'_\lambda}^2$ as
\begin{equation}
k = \frac{\sigma_2^2}{\sigma_1^2} \frac{m}{n} = \frac{a_1}{\sigma_1^2 b_1} / \frac{a_2}{\sigma_2^2 b_2},
\end{equation}
i.e., the weight of each line can be expressed as
\begin{equation}
w_i = \frac{a_i}{\sigma_i^2 b_i}.
\end{equation}

Therefore, we adopt Equation (B10) as the weight of each line to merge all data to a co-added spectrum.

\begin{figure}[!t]
\center{
\includegraphics[angle=0,scale=0.50]{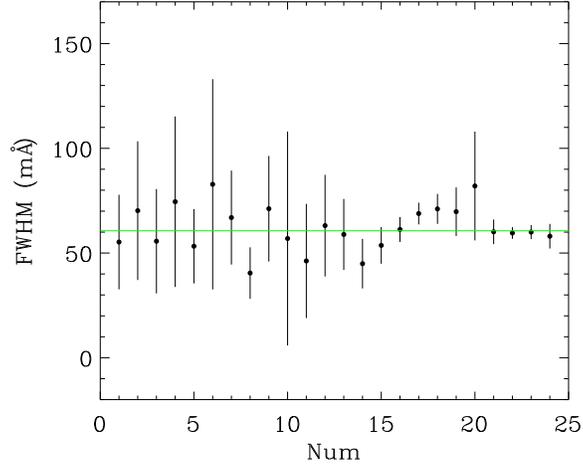}}
\caption{The distribution of the observed Gaussian broadening $b$ of the \oi~K$\alpha$ line.
         Here only the line fitting results with $\sigma_b < 30$ m\AA\ are plotted, which are consistent
         with the same value.}
\label{Fig:14}
\end{figure}


\begin{thebibliography}{}
\bibitem[Behar \& Netzer(2002)]{2002ApJ...570.165} Behar, E., \& Netzer, H.\ 2002, \apj, 570, 165
\bibitem[Bowen et al.(2008)]{2008ApJS...176.59} Bowen, D.~V., et al.\ 2008, \apjs, 176, 59
\bibitem[Cabot et al.(2013)]{2013MNRAS...431.511} Cabot S. H., Wang, D. Q., \& Yao, Y.\ 2013, \mnras, 431, 511
\bibitem[Dewey et al.(2008)]{2008ApJ...676.L131} Dewey, D., Zhekov, S.~A., McCray, R., \& Canizares, C.~R.\ 2008, \apj, 676, L131
\bibitem[Fang et al.(2003)]{2003ApJ...586.L49} Fang, T.~T., Sembach, K.~R., Canizares, C.~R.\ 2003, \apj, 586, L49
\bibitem[Galloway et al.(2008)]{2008ApJ...179.360} Galloway, D. K., Muno, M. P., Hartman, J. M., Psaltis, D., \& Chakrabarty, D.\ 2008a, \apjs, 179, 360
\bibitem[Garc\'ia et al.(2005)]{gar05} Garc\'ia, J., Mendoza, C., Bautista, M.~A., Gorczyca, T.~W., Kallman, T.~R., Palmeri, P.\ 2005, \apj, 158, 68
\bibitem[Gorczyca(2000)]{gor00} Gorczyca, T. 2000, Phys. Rev. A., 61, 024702
\bibitem[Gorczyca \& McLaughlin(2005)]{gor05} Gorczyca, T., \& McLaughlin, B.\ 2005, BAPS, 50, 39
\bibitem[Iaria et al.(2004)]{2004ApJ...600.358} Iaria, R., et al.\ 2004, \apj, 600, 358
\bibitem[Juett et al.(2004)]{2004ApJ...612.308} Juett, A.~M., Schulz, N.~S., \& Chakrabarty, D.\ 2004, \apj, 612, 308
\bibitem[Juett et al.(2006)]{2006ApJ...648.1066} Juett, A.~M., Schulz, N.~S., Chakrabarty, D., \& Gorczyca, T.~W.\ 2006, \apj, 648, 1066
\bibitem[Juett \& Chakrabarty(2006)]{2006ApJ...646.493} Juett, A.~M., \& Chakrabarty, D.\ 2006, \apj, 646, 493
\bibitem[Kalberla et al.(2005)]{2005A&A...440.775} Kalberla, P.~M.~W., et al.\ 2005, \aa, 440, 775
\bibitem[Kallman et al.(2004)]{2004ApJS...155.675} Kallman, T.~R., Palmeri, P., Bautista, M.~A., Mendoza, C., \& Krolik, J.~H.\ 2004, \apjs, 155, 675
\bibitem[Kong et al.(2006)]{2006MNRAS...368.781} Kong, A.~K.~H., Charles, P.~A., Homer, L., Kuulkers, E., \& O'Donoghue, D.\ 2006, \mnras, 368, 781
\bibitem[Kuulkers et al.(2010)]{2010A&A...514.A65} Kuulkers, E., et al.\ 2010, \aa, 514, A65
\bibitem[Landman et al.(1982)]{1982ApJ...261.732} Landman, D.~A., Robert, R.~D., Tanigawa, G.\ 1982, \apj, 261, 732
\bibitem[Lee et al.(2001)]{2001ApJ...554.L13} Lee, J.~C., et al.\ 2001, \apj, 554, L13
\bibitem[Lenz \& Ayres (1992)]{1992PASP...104.1104} Lenz, D.~D., \& Ayres, T.~R.\ 1992, PASP, 104, 1104
\bibitem[Miller et al.(2004)]{2004ApJ...601.450} Miller, J.~M., et al.\ 2004, \apj, 601, 450
\bibitem[Miller et al.(2006)]{2006Nature...441.953} Miller, J.~M., et al.\ 2006, \nat, 441, 953
\bibitem[Morrison \& McCammon (1983)]{MM83} Morrison, R., \& McCammon, D.\ 1983, \apj, 270, 119
\bibitem[Nicastro et al.(2005)]{2005ApJ...629.700} Nicastro, F., et al.\ 2005, \apj, 629, 700
\bibitem[Rand(1997)]{1997ApJ...474.129} Rand, R.~J.\ 1997, \apj, 474, 129
\bibitem[Rand(2000)]{2000ApJ...537.L13} Rand, R.~J.\ 2000, \apj, 537, L13
\bibitem[Raassen et al.(2003)]{2003A&A...400.671} Raassen, A.~J.~J., Ness, J.~U., Mewe, R., van der Meer, R.~L.~J., Burwitz, V., \& Kaastra, J.~S.\ 2003, \aa, 400, 671
\bibitem[Sala et al.(2007)]{2007Ap&SS...309.315} Sala, G., Greiner, J., Bottacini, E., \& Haberl, F.\ 2007, Ap\&SS, 309, 315
\bibitem[Schattenburg & Canizares(1986)]{1986ApJ...301.759} Schattenburg, M.~L., \& Canizares, C.~R.\ 1986, \apj, 301, 759
\bibitem[Schulz et al.(2002)]{2002ApJ...565.1141} Schulz, N. S., et al.\ 2002, \apj, 565, 1141
\bibitem[Smith et al.(2001)]{2001ApJ...556.L91} Smith, R.~K., Brickhouse, N.~S., Liedahl, D.~A., \& Raymond, J.~C.\ 2001, ApJ, 556, L91
\bibitem[Sparke \& Gallagher (2000)]{spa00} Sparke, L. S., \& Gallagher, J. S. 2000, Galaxies in the Universe: an Introduction, Cambridge University Press
\bibitem[Steenbrugge et al.(2003)]{2003A&A...402.477} Steenbrugge, K.~C., Kaastra, J.~S., de Vries, C.~P., \& Edelson, R.\ 2003, \aa, 402, 477
\bibitem[Stolte et al.(1997)]{1997J.Phys.B...30.4489} Stolte, W.~C., et al.\ 1997, J. Phys. B, 30, 4489
\bibitem[Verner et al.(1996)]{1996ADNDT...64.1} Verner, D.~A., Verner, E.~M., \& Ferland, G.~J.\ 1996, ADNDT, 64, 1
\bibitem[Yao \& Wang(2005)]{2005ApJ...624.751} Yao, Y., \& Wang, Q,~D.\ 2005, \apj, 624, 751
\bibitem[Yao et al.(2008)]{2008ApJ...672.L21} Yao, Y., Nowak, M.~A., Wang, Q.~D., Schulz, N.~S., \& Canizares, C.~R.\ 2008, \apj, 672, L21
\bibitem[Yao et al.(2009)]{2009ApJ...696.1418} Yao, Y., Schulz, N.~S., Gu, M.~F., Nowak, M.~A. \& Canizares, C.~R.\ 2009, \apj, 696, 1418

\end{thebibliography}
\end{document}